\documentclass[%
 reprint,
 amsmath,amssymb,
 aps,
]{revtex4-1}
\usepackage{graphicx}
\usepackage{dcolumn}
\usepackage{bm}
\usepackage{graphicx}
\usepackage{dcolumn}
\usepackage{bm}
\usepackage{bigints}
\usepackage{color}
\usepackage{amsmath,amssymb,graphicx}
\usepackage{float}
\usepackage{braket}
\usepackage{tabularx}
\usepackage{tabulary}
\usepackage{tabu}
\usepackage{booktabs}
\usepackage{textcomp}

\newcolumntype{C}{>{\centering\arraybackslash}X}

\begin{document}
\preprint{APS/123-QED}
\title{Heat mitigation of a core/cladding Yb-doped fiber amplifier using anti-Stokes fluorescence cooling}

\author{Esmaeil Mobini$^{1,2}$}%
\author{Mostafa Peysokhan$^{1,2}$}%
\author{Arash Mafi$^{1,2,}$}%

\email{mafi@unm.edu}
\affiliation{$^1$Department of Physics \& Astronomy, University of New Mexico, Albuquerque, NM 87131, USA \\
             $^2$Center for High Technology Materials, University of New Mexico, Albuquerque, NM 87106, USA} 

\begin{abstract}
A Core/Cladding Yb-doped fiber amplifier configuration is presented that relies on anti-Stokes fluorescence cooling for effective heat mitigation in high-power operation. In the proposed design, the inner cladding of the double clad fiber is doped with the same ion as in the core; therefore, the excess heat generated from the background absorption is removed by the anti-Stokes fluorescence in both core and inner cladding. We consider both silica and ZBLAN glasses for the host material in the Core/Cladding Yb-doped fiber amplifier. The model incorporates the spatial profiles of the signal and pump intensities, as well as the amplified spontaneous emission. The total linear heat density and temperature distribution in the fiber amplifier are calculated. The results show that the anti-Stokes fluorescence cooling in the Core/Cladding Yb-doped configuration can mitigate the generated heat effectively in high-power operation, which obviates the need for an external cooling system.   
\end{abstract}

\maketitle

\section{Introduction}
Over the past decades, high-power fiber lasers and amplifiers have been extensively under research to achieve higher output powers. However, temperature rise in the core of the fiber lasers and amplifiers has been a hindrance to an acceptable stability and efficiency in high power operation~\cite{brown2001thermal,Richardson,ward2012origin,Dawson:08}. Strides have been constantly made to solve the perennial excess heat issue in the high-power fiber lasers and amplifiers using different methods such as liquid-forced cooling~\cite{Dawson:08}. However, the need for an ever more effective heat mitigation seems to be pivotal to achieving a higher efficiency in the design of fiber lasers and amplifiers, especially in high-power operation. 

Radiation-Balancing is a viable technique that has been introduced for effective heat mitigation in high-power lasers by S.~Bowman~\cite{bowman1999lasers,bowman2010minimizing,yang2019radiation}.
The radiation balanced lasers (RBL) rely on solid-state laser cooling: the gain medium is pumped at a wavelength of $\lambda_{p}$ which is longer than the mean fluorescence wavelength ($\lambda_{f}$) of the doped-material to ensure anti-Stokes fluorescence cooling~\cite{Pringsheim1929,epstein1995observation,seletskiy2010laser}. In a laser, the signal wavelength ($\lambda_{s}$) is longer than $\lambda_{p}$ and naturally $\lambda_{f}$. As a subtle balance condition for RBL, the excess heat generated by the quantum defect and background absorption can be extracted by the anti-Stokes fluorescence cooling; therefore, the net heat can be zero. However, because the mean fluorescence wavelength of the ion-doped materials lies in the tail of the absorption cross-section spectrum, one should consequently expect some differences in the design of the radiation-balanced fiber lasers or amplifiers compared to the conventional Double Clad (DC) fiber lasers 
and amplifiers~\cite{nemova2009athermal,8423184,Mobini:188}.

In our previous study~\cite{Mobini:188}, we explored the heat mitigation by anti-Stokes fluorescence cooling in a typical Yb-doped silica DC fiber amplifier.
The calculations show that the anti-Stokes fluorescence cooling can mitigate the generated heat effectively in a typical DC fiber amplifier for powers up to a few tens of Watts. But at higher powers, 
e.g., hundreds of Watts and more, the background absorption of the inner cladding becomes the dominant factor contributing to the heat generation in the DC fiber amplifiers. The heat generated due to the background absorption in the inner cladding outweighs significantly the maximum heat extraction that can be delivered by the anti-Stokes fluorescence cooling inside the core; therefore, for the high power operation of a typical DC fiber amplifier, the anti-Stokes fluorescence cooling fails to mitigate the generated heat effectively. Hence, a new configuration should be introduced in which the heat extraction by the anti-Stokes fluorescence cooling is large enough to counterbalance the significant heat load from the background absorption generated in the inner cladding in high-power operation. 

A viable method to cancel out the excess heat in the inner cladding is to dope the inner cladding with the same ions as in the core. In a DC fiber laser or amplifier, the inner cladding is also pumped; therefore, the doped inner cladding can also contribute to the cooling process. Because the inner cladding area is much larger than the core area, a properly doped inner cladding can increase the cooling power sufficiently to counterbalance the heat generation in the high-power operation. In our analysis, both the core and inner cladding are doped with the same Yb ion; therefore, we refer to this design as the ``Core/Cladding (C/C) Yb-doped fiber amplifier''. 

A similar model was also suggested by R. Kashyap et al.~\cite{Nemova:09}, in which the cladding of a ZBLAN multimode fiber amplifier was used as an integrated optical cooler to offset the excess heat generated in the core. In their model, the host material was ZBLAN glass and the core and cladding were doped with Yb and Tm ions, respectively. The presence of two different ions necessitated two different pump sources. In their model, in order to satisfy the RBL condition at every point along the fiber, the dopant density of the ions were proposed to vary in the longitudinal direction. Our single ion composition makes the implementation of our proposed design easier. Moreover, we include the background absorption in our analysis~\cite{Mobini:188}, which in practice is the dominant heat source in high-power radiation-balanced fiber lasers and amplifiers.
 
The goal of the present work is to explore the heat mitigation via anti-Stokes fluorescence cooling in the C/C Yb-doped fiber amplifier using a comprehensive thermal modeling that includes the main factors that contribute to the heat generation. The detailed formalism considers the spatial profiles of both signal and pump 
intensities across the C/C Yb-doped configuration, the presence of the amplified spontaneous emission (ASE) as the source of anti-Stokes fluorescence cooling, and 
the temperature variation across and along the C/C Yb-doped configuration. The study first starts with the calculations of the right- and left- moving pump and signals powers along the C/C Yb-doped fiber amplifier and takes into account the contributions of the ASE consistently. Using the calculated pump and signals powers along the C/C Yb-doped fiber amplifier, the contributions of the pump, signals and anti-Stokes fluorescence cooling to the heat density ($Q$) are calculated and finally the temperature distribution along and across the C/C Yb-doped amplifier is obtained. The results show that the anti-Stokes fluorescence cooling can mitigate the generated heat effectively. 

As the host material of the C/C Yb-doped fiber amplifier, we consider both silica and ZBLAN glasses. Silica glass has been proven to 
be a reliable host material for the fiber gain medium in the high power operation due to its strong chemical and physical stability and its high optical power damage 
threshold~\cite{Dawson:08}. However, so far no experimental result on solid-state laser cooling on the Yb-doped silica glass has been published. Hence, we recently performed a detailed investigation and raised the possibility of the solid-state laser cooling in Yb-doped silica glass~\cite{Mobini-PhysRevApplied.11.014066}. The results in Ref.~\cite{Mobini-PhysRevApplied.11.014066}, which were based on the spectroscopic investigations over a range of different temperatures have revealed that there is no a priori fundamental reason to prevent solid-state laser cooling in a high-purity Yb-doped silica glass. Therefore, Yb-doped silica glass is adopted as a potentially cooling-grade host material in the present study. The second host material that is implemented for this study is the ZBLAN glass, which has been known as a cooling-grade material. Although ZBLAN glass in terms of chemical and physical stability is not as reliable as the silica glass, its long successful record for the solid-state laser cooling makes it a good candidate for the C/C Yb-doped configuration~\cite{epstein1995observation,gosnell1999laser}.

One of the inherent difficulties of the RBL technique is that the pump wavelength has to be in the tail of the absorption cross-section to 
exceed the mean fluorescence wavelength. Therefore, the signal amplification drops appreciably due to low absorption cross-section in the RBLs~\cite{Mobini:188,yang2019radiation}. It must be noted that the goal of the present work is not to deal with the low signal amplification issue in the RBL, which lies beyond the scope of the study; rather, the main idea is to implement the Yb-doped silica 
and ZBLAN glasses as the host materials for the suggested C/C Yb-doped configuration.        

Finally, we discuss the impact of the inner cladding diameter on the total linear heat density (LHD) and signal efficiency ($\eta_{s}$) of the C/C Yb-doped configuration. The results show that one can sacrifice a better signal efficiency (a higher signal amplification) in favor of having more effective heat mitigation in the suggested configuration and vice versa. In other words, in the suggested configuration, the signal efficiency and heat mitigation through anti-Stokes fluorescence cooling are incompatible with each other. It is also important to mention that the intention of the paper is not to find the optimum parameters for the suggested design; rather, its purpose is to show a few examples where the proposed configuration can effectively offset the generated heat as a proof of concept.   

\section{Basic Formalism}
\begin{figure}[!h]
\centering
    \includegraphics[width=1.3 in]{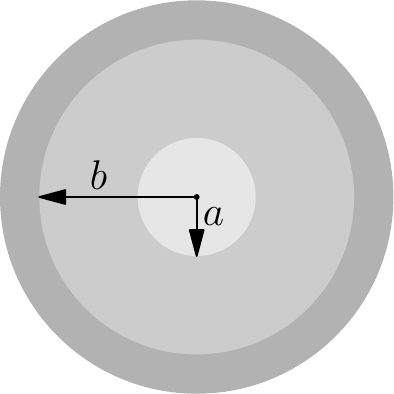}
\caption{Schematic of a C/C ion-doped fiber amplifier.}
\label{Fig:Cross-sectional-fiber}
\end{figure}

The optical fiber considered here is very similar to a typical DC fiber, except for its inner cladding, which is doped with the same ion as in the core. The 
cross sectional schematic of the proposed C/C ion-doped fiber is shown in Fig.~\ref{Fig:Cross-sectional-fiber}, where $a$ and $b$ represent the
core and inner cladding radii, respectively.

The total dopant distribution ($N_{0}$) in the core and the inner cladding is defined by 
\begin{align}
N_{0}(\rho)&=
\begin{cases}
N_{01}, &\text{$\rho \le a $}\\
N_{02}, &\text{$a~<\rho \le b $}\\
\end{cases},
\label{Eq:dopant-dist}
\end{align}
where $N_{01}$ and $N_{02}$ are the dopant densities in the core and inner cladding, respectively, and $\rho$ represents the radial coordinate.

In general, the core and inner cladding are doped with different ion densities. The dopant density is known to affect the fluorescence lifetime ($\tau_{F}$)~\cite{auzel2003radiation,
Mobini-PhysRevApplied.11.014066}. Therefore, we differentiate between the fluorescence lifetimes of the core and inner cladding by: 
\begin{align}
\tau_{F}(\rho)&=
\begin{cases}
\eta_{q1}\tau_{r}, &\text{$\rho \le a $}\\
\eta_{q2}\tau_{r}, &\text{$a~<\rho \le b $}\\
\end{cases},
\end{align}
where $\eta_{q1}$ and $\eta_{q2}$ are the internal quantum efficiencies of the core and the inner cladding of the fiber, respectively, 
and $\tau_{r}$ is the radiative lifetime. Here due to the fact that both the core and the inner cladding are doped with the same dopant ion, 
we have assumed that the radiative lifetime, $\tau_{r}$, is the same in the core and the inner cladding.
\subsection{Pump power propagation}
To calculate the pump and signal (laser and ASE) intensities in the core and the inner cladding of the fiber, we follow the formalism introduced in Ref.~\cite{Mobini:188}.
Unlike a typical DC fiber amplifier in which the pump power is only attenuated by the background absorption in the inner cladding, here in the suggested C/C ion-doped fiber amplifier, 
the pump power is attenuated by the background absorption and resonant absorption originating from the dopant in the inner cladding. Therefore,
the impact of the dopant in the inner cladding should be taken into account in the modeling. 

In the following, we refer to the pump frequency (wavelength) as $\nu_p(\lambda_p)$. The signal 
spectrum (laser and ASE) is sliced into $n$ adjacent segments, where $\delta\lambda$ is the bandwidth 
for each segment. The signal frequencies and wavelengths are $\nu_j(\lambda_j)$, where $j=1,\cdots,n$.
We make a convenient assumption that the signals propagate in a Gaussian beam profile 
with the mode field diameter ($w$), which is independent of $\lambda_j$~\cite{8423184,Mobini:188}: 
\begin{align}
g_{w}(\rho)=e^{-2(\frac{\rho}{w})^2}.
\label{Eq:Gauss-beam}
\end{align}
Hence, the signal intensity for each spectral slice ($I_{j}(\rho,z)$) takes the form of
\begin{align}
f_{w}(\rho)&=2\frac{g_{w}(\rho)}{\pi w^2},\quad \quad I_{j}(\rho,z)=f_{w}(\rho) P_{j}(z), 
\label{Eq:signal-int}
\end{align}
where $\int_{0}^{\infty}f_{w}(\rho)~2\pi~\rho d\rho\,=\,1$ and $P_{j}(z)$ represents the signal power along the fiber. The fiber is assumed to be cylindrically symmetric;
therefore, $\int_{0}^{2\pi}I_{j}(\rho,\phi,z) d\phi\,=\,2\pi~I_{j}(\rho,z)$, where $\phi$ represents the azimuthal angle.
Similar to a DC fiber amplifier, for the C/C ion-doped configuration we assume that the pump intensity, $I_{p}(\rho,z)$, is uniformly distributed across the 
fiber amplifier as described by
\begin{align}
I_{p}(\rho,z)=\frac{P_{p}(z)}{\pi b^2},
\label{Eq:Pump-int}
\end{align}
where $P_{p}(z)$ represents the pump power along the fiber amplifier.

Using derived equations in Ref.~\cite{Mobini:188}, the right- ($I^{+}_{p}$) and left- ($I^{-}_{p}$) moving pump intensities along the fiber 
can be written as   
\begin{align}
\label{Eq:pump-power1}
\pm \frac{dI^{\pm}_{p}}{dz}&=-N_{0}\sigma_{p}^{a}\frac{1+\sum\limits_{j=1}^n \gamma_{j} \tilde{i}_{j}^{\pm}}{1+\tilde{i}_{p}^{\pm}+\sum\limits_{j=1}^n 
\tilde{i}_{j}^{\pm}} I^{\pm}_{p}-\alpha_{b} I^{\pm}_{p},
\end{align}
where the implemented identities are defined as 
\begin{align}
&i_p^{\pm}=\frac{I_p^{\pm}}{I_p^{sat}},\qquad &&i_j^{\pm}=\frac{I_{j}^{\pm}}{I_{j}^{sat}},\nonumber\\
&\tilde{i}_{p}^{\pm}= {i}_{p}^{+}+{i}_{p}^{-},\qquad &&\tilde{i}_{j}^{\pm}= {i}_{j}^{+}+{i}_{j}^{-},\nonumber\\
&\tilde{p}_{p}^{\pm}= {p}_{p}^{+}+{p}_{p}^{-},\qquad &&\tilde{p}_{j}^{\pm}= {p}_{j}^{+}+{p}_{j}^{-},\nonumber\\
&p_{j}^{\pm}=\frac{{P}_{j}^{\pm}}{{P}_{j}^{sat}},\qquad && p_{p}^{\pm}=\frac{{P}_{p}^{\pm}}{{P}_{p}^{sat}},\nonumber\\ 
&P_{p}^{sat}=\overline{I_{p}^{sat}} \pi b^2,\qquad && P_{j}^{sat}=\overline{I_{j}^{sat}} A_{eff} ,\quad A_{eff}=\frac{\pi w^2}{2},
\end{align}
and $\alpha_{b}$ is the background absorption coefficient.

Because as indicated earlier, the fluorescence lifetime is generally different in the core and the inner cladding, the saturation intensity naturally becomes
different in the core and the inner cladding as follows:
\begin{align}
I_{p}^{sat}(\rho)&=\overline{I_{p}^{sat}}\times
\begin{cases}
\eta_{q1}^{-1}, &\text{$\rho \le a $}\\
\eta_{q2}^{-1}, &\text{$a~<\rho \le b $}\\
\end{cases},\nonumber\\
I_{j}^{sat}(\rho)&=\overline{I_{j}^{sat}}\times
\begin{cases}
\eta_{q1}^{-1}, &\text{$\rho \le a $}\\
\eta_{q2}^{-1}, &\text{$a~<\rho \le b $}\\
\end{cases},\nonumber\\
\overline{I_{p}^{sat}}&=\frac{hc\beta_{p}}{\lambda_{p}\tau_{r}\sigma_{p}^{a}},\quad \quad \overline{I_{j}^{sat}}=\frac{hc \beta_{j} }{\lambda_{j}\tau_{r}\sigma_{j}^{a}},\nonumber\\
\beta_{p}&=\frac{\sigma_{p}^{a}}{\sigma_{p}^{a}+\sigma_{p}^{e}},\quad \quad \beta_{j}=\frac{\sigma_{j}^{a}}{\sigma_{j}^{a}+\sigma_{j}^{e}},\quad \quad
\gamma_{j}=1-\frac{\beta_{j}}{\beta_{p}}.
\end{align}
The emission and absorption cross-sections at the corresponding signal slices are $\sigma_{j}^{e}\,=\,\sigma^{e}(\lambda_{j})$ 
and $\sigma_{j}^{a}\,=\,\sigma^{a}(\lambda_{j})$, respectively, and the emission and absorption cross-sections at the pump wavelength 
are given as $\sigma_{p}^{e}\,=\,\sigma^{e}(\lambda_{p})$ and $\sigma_{p}^{a}\,=\,\sigma^{a}(\lambda_{p})$, respectively.

In order to obtain the longitudinal evolution of the right- and left- moving pump powers, we integrate the intensities over the entire cross-section of the fiber: 
$\int^{\infty}_{0}I_{p}^{\pm}(\rho,z)~2\pi \rho d\rho=P_{p}^{\pm}$. Therefore, the differential equations that describe the right- and left- propagating pump powers in the fiber 
are obtained and can be described by (See Appendix~A for details)
\begin{align}
&\pm\frac{dP_p^{\pm}}{dz}=-\alpha_{b}P^{\pm}_p
-\sigma_{p}^{a}N_{01}\Big(\frac{\mathcal{A}}{\mathcal{C}_{1}} \Gamma_{p}+\Gamma(\frac{\mathcal{A}\mathcal{D}_{1}-\mathcal{B}_{1}\mathcal{C}_{1}}{\mathcal{C}_{1}\mathcal{D}_{1}})\nonumber \\
&\times \ln\big[1-\eta_{a}\frac{\mathcal{D}_{1}}{\mathcal{C}_{1}+\mathcal{D}_{1}}\big]-\sigma_{p}^{a}N_{02}\Big(\frac{\mathcal{A}}{\mathcal{C}_{2}} (1-\Gamma_{p}) \nonumber \\
&+\Gamma(\frac{\mathcal{A}\mathcal{D}_{2}-\mathcal{B}_{2}\mathcal{C}_{2}}{\mathcal{C}_{2}\mathcal{D}_{2}})
\times \ln\big[\frac{\mathcal{C}_{2}+(1-\eta_{b})\mathcal{D}_{2}}{\mathcal{C}_{2}+(1-\eta_{a})\mathcal{D}_{2}}\big]\Big)P^{\pm}_{p},
\label{Eq:pump-power4}
\end{align}
where 
\begin{align}
&\mathcal{A}=1,\quad 
\mathcal{B}=\sum\limits_{j=1}^n \gamma_{j} \tilde{p}_{j}^{\pm},\quad 
\mathcal{C}=1+\tilde{p}_{p}^{\pm},\quad
\mathcal{D}=\sum\limits_{j=1}^n\tilde{p}_{j}^{\pm},\nonumber\\
&\mathcal{B}_{i}=\eta_{qi}\mathcal{B},\qquad
\mathcal{C}_{i}=(1-\eta_{qi})+\eta_{qi}\mathcal{C},\qquad
\mathcal{D}_{i}=\eta_{qi}\mathcal{D},\nonumber\\
&\eta_a=1-\exp{\big(-2a^2/w^2\big)}, \quad \eta_b=1-\exp{\big(-2b^2/w^2\big)}.
\end{align}
Here $i\in\{1,2\}$, where $i=1$ refers to the core and $i=2$ refers to the inner cladding. 
$\Gamma_{p}=(a/b)^2$ and $\Gamma$=\,A\,$_{eff}/\pi b^2$  are the pump overlap-factor and the signal overlap-factor, respectively.
\subsection{Signal power propagation}
Similar to the pump propagation, the differential equations describing the left- and right-moving signal intensities along the fiber can be described by~\cite{Mobini:188}:
\begin{align}
\pm \frac{dI^{\pm}_{j}}{dz}&=\frac{\beta_{p}\tilde{i}_{p}^{\pm}+\sum\limits_{k=1}^n \beta_{k} \tilde{i}_{k}^{\pm}}{1+\tilde{i}_{p}^{\pm}+\sum\limits_{k=1}^n 
\tilde{i}_{k}^{\pm}} N_{0}\Big((\sigma_{j}^{a}+\sigma_{j}^{e})I_{j}^{\pm}+\sigma_{j}^{e}\Pi_{j}f_{\omega}\Big)\nonumber\\
&-N_{0} \sigma_{j}^{a}I_{j}^{\pm} -\alpha_{b} I^{\pm}_{j},
\label{Eq:signal-intensity}
\end{align}
where $\Pi_{j}=2 h c^2 \delta\lambda/\lambda_{j}^{3}$ represents the local noise source for ASE. 

Integrating Eq.~\ref{Eq:signal-intensity} over the entire dopant area, including the inner cladding, the differential equations for the right- and left-moving signal powers 
along the fiber amplifier are obtained (See Appendix~B for details):
\begin{align}
&\pm \frac{dP_{j}^{\pm}}{dz}=-\alpha_{b}\eta_{b}P_{j}^{\pm}+N_{01}\Big((\sigma_{j}^{a}+\sigma_{j}^{e})P_{j}^{\pm}+\sigma_{j}^{e}\Pi_{j}\Big) \times \nonumber\\
& \Big( \frac{\overline{\mathcal{B}}_{1}}{\mathcal{D}_{1}}\eta_{a}-(\frac{\overline{\mathcal{A}}_{1}\mathcal{D}_{1}-\overline{\mathcal{B}}_{1}\mathcal{C}_{1}}{\mathcal{D}_{1}^2}) 
\ln\big(1-\eta_{a}\frac{\mathcal{D}_{1}}{\mathcal{C}_{1}+\mathcal{D}_{1}}\big)\Big)\nonumber\\
&-N_{01} \sigma_{j}^{a}P_{j}^{\pm}\eta_{a}+N_{02}\Big((\sigma_{j}^{a}+\sigma_{j}^{e})P_{j}^{\pm}+\sigma_{j}^{e}\Pi_{j}\Big) \times\nonumber\\
& \Big(\frac{\overline{\mathcal{B}_{2}}}{\mathcal{D}_{2}}(\eta_{b}-\eta_{a})-(\frac{\overline{\mathcal{A}}_{2}\mathcal{D}_{2}-\overline{\mathcal{B}}_{2}\mathcal{C}_{2}}{\mathcal{D}_{2}^2})
\ln\big(\frac{\mathcal{C}_{2}+\mathcal{D}_{2}(1-\eta_{b})}{\mathcal{C}_{2}+\mathcal{D}_{2}(1-\eta_{a})}\big)\Big)\nonumber\\
&-N_{02} \sigma_{j}^{a}P_{j}^{\pm}(\eta_{b}-\eta_{a}),
\label{Eq:signal-power-2ndterm}
\end{align}
where 
\begin{align}
\mathcal{\overline{A}}=\beta_{p}\tilde{p}_{p}^{\pm}, \quad
\mathcal{\overline{A}}_{i}=\eta_{qi}\mathcal{\overline{A}}, \quad 
\mathcal{\overline{B}}=\sum\limits_{j=1}^n \beta_{j} \tilde{p}_{j}^{\pm},\quad 
\mathcal{\overline{B}}_{i}=\eta_{qi}\mathcal{\overline{B}}.
\end{align}
\subsection{Fluorescence emission (anti-Stokes cooling)}
In order to calculate the contribution of anti-Stokes fluorescence to the heat generation, we need to obtain the average energy that escapes from the 
fiber amplifier via fluorescence emission. Here we assume that the total ion density in the excited state decays over the radiative lifetime with an average energy of $E_{f}=hc/\lambda_{f}$ for each decaying excited ion. This process can be quantitatively described by     
\begin{align}
\frac{dI_{f}}{dz}=\frac{hc}{\lambda_{f}\tau_{r}(\rho)}N_{2}(\rho),
\label{Eq:FluorEm0}
\end{align}
where $N_2$ represents the total ion density in the excited state and can be described by~\cite{Mobini:188}
\begin{align}
N_{2}(\rho)=\frac{\beta_{p}\tilde{p}_{p}^{\pm}+g_{w}(\rho)\sum\limits_{j=1}^n \beta_{j} \tilde{p}_{j}^{\pm}}{1+\tilde{p}_{p}^{\pm}+g_{w}(\rho)\sum\limits_{j=1}^n \tilde{p}_{j}^{\pm}} N_{0}(\rho).
\label{Eq:Upperlevel-Pop}
\end{align}
Integrating Eq.~\ref{Eq:FluorEm0} over the entire dopant area, the fluorescence linear power can be obtained (See Appendix D for details). 
The fluorescence linear power along the C/C ion-doped fiber amplifier can be described by
\begin{align}
&\frac{dP_{f}}{dz}=\frac{hc}{\lambda_{f}\tau_{r}} N_{01}\Big(\frac{\overline{\mathcal{A}}_{1}}{\mathcal{C}_{1}}\pi a^2+\frac{\pi w^2}{2}(\frac{\overline{\mathcal{A}}_{1}\mathcal{D}_{1}-\overline{\mathcal{B}}_{1}\mathcal{C}_{1}}{\mathcal{C}_{1}\mathcal{D}_{1}})\nonumber\\
&\times \ln\big[1-\eta_{a}\frac{\mathcal{D}_{1}}{\mathcal{D}_{1}+\mathcal{C}_{1}}\big]\Big)\nonumber\\
&+\frac{hc}{\lambda_{f}\tau_{r}}N_{02}\Big(\frac{\overline{\mathcal{A}}_{2}}{\mathcal{C}_{2}}(\pi b^2-\pi a^2)+\frac{\pi w^2}{2}(\frac{\overline{\mathcal{A}}_{2}\mathcal{D}_{2}-\overline{\mathcal{B}}_{2}\mathcal{C}_{2}}{\mathcal{C}_{2}\mathcal{D}_{2}})\nonumber\\
&\times \ln\big[\frac{\mathcal{C}_{2}+(1-\eta_{b})\mathcal{D}_{2}}{\mathcal{C}_{2}+(1-\eta_{a})\mathcal{D}_{2}}\big]\Big).
\label{Eq:FluorEm-1}
\end{align}

Here we have assumed that the entire fluorescence emission, except a small fraction that is guided through the fiber and seeds the ASE, escapes from the fiber amplifier, 
which is an acceptable assumption for the fiber due to its small cross-section~\cite{ruan2006enhanced}.    
\section{Heat generation and extraction}
We now have all the necessary ingredients to calculate the total LHD and consequently the temperature distribution of the fiber amplifier by solving the steady-state heat equation. 
Here, we have broken the total LHD ($q$) into separate components as~\cite{Mobini:188} 
\begin{align}
\label{Eq:LinearHeat0}
\nonumber
q(\tilde{P}_p^{\pm},\tilde{P}_j^{\pm})&=q_{p}(\tilde{P}_p^{\pm},\tilde{P}_j^{\pm})+q_{s}(\tilde{P}_p^{\pm},\tilde{P}_j^{\pm})\\
&+q_{f}(\tilde{P}_p^{\pm},\tilde{P}_j^{\pm})-q_{b,s}(\tilde{P}_p^{\pm},\tilde{P}_j^{\pm}).
\end{align}
$q_{p}$ and $q_{s}$ denote the contributions of the pump and signal powers. The subtraction of $q_{b,s}$ signifies the scattering part of the background absorption coefficient 
that does not contribute to the heat generation. In other words, the background absorption coefficient consists of an absorptive and a scattering part, $\alpha_{b,a}$ and $\alpha_{b,s}$, 
respectively ($\alpha_b=\alpha_{b,a}+\alpha_{b,s}$)~\cite{Peysokhan:18}--the scattering part does not lead to the heat generation. 
$q_{f}$ represents the anti-Stokes fluorescence cooling.
\begin{align}
\label{Eq:LinearHeat}
&q_{p}=-\frac{dP_{p}^{+}}{dz}+\frac{dP_{p}^{-}}{dz},\\ 
&q_{s}=\sum\limits_{j=1}^n\big(-\frac{dP_{j}^{+}}{dz}+\frac{dP_{j}^{-}}{dz}\big),\nonumber\\
&q_{f}=-\frac{dP_{f}}{dz}+\int_{0}^{b} 2\pi\rho d\rho \big( 2\sum\limits_{j=1}^n N_2 \sigma_{j}^{e}\Pi_{j} f_{w}\big),\nonumber\\
&q_{b,s}=
\begin{cases}
\alpha_{b,s}\big(\Gamma_{P}\tilde{P}_{p}^{\pm}+\eta_{a}\sum\limits_{j=1}^n\tilde{P}_{j}^{\pm}\big), &\text{$\rho \le a $}\\
\alpha_{b,s}\big((1-\Gamma_{p})\tilde{P}_{p}^{\pm}+(\eta_{b}-\eta_{a})\sum\limits_{j=1}^n\tilde{P}_{j}^{\pm}\big), &\text{$a <\rho \le b $}\nonumber\\
\end{cases}.
\end{align}

In order to obtain the temperature distribution in the amplifier, we need to solve the steady-state heat differential equation, where the source is the volume heat density, rather than the LHD ($q$)~\cite{brown2001thermal}. The volume heat density $Q$ for the core and the inner cladding areas are given by:
\begin{align}
&Q_{co}(\tilde{P}_p^{\pm},\tilde{P}_j^{\pm})=\frac{1}{\pi a^2} q(\tilde{P}_p^{\pm},\tilde{P}_j^{\pm}), &\rho \le a,\label{Eq:VHD}\nonumber\\
&Q_{inc}(\tilde{P}_p^{\pm},\tilde{P}_j^{\pm})=\frac{1}{(1-\Gamma_{p})}\frac{1}{\pi b^2}q(\tilde{P}_p^{\pm},\tilde{P}_j^{\pm}), & a <\rho \le b,
\end{align}
where $Q_{co}$ and $Q_{inc}$ denote the heat densities in the core and the inner cladding of the fiber, respectively.

We have assumed that the volume heat density inside the core and the inner cladding is uniform. Inserting the volume heat densities from Eq.~\ref{Eq:VHD}
into the heat equation allows us to find the temperature distribution in the core ($T_{co}$) and the inner cladding ($T_{inc}$) at each position along the 
fiber (See Appendix D for details)~\cite{brown2001thermal}.
\section{Core/cladding Yb-doped fiber amplifier}
\begin{figure}[!h]
\centering
    \includegraphics[width=3.5 in]{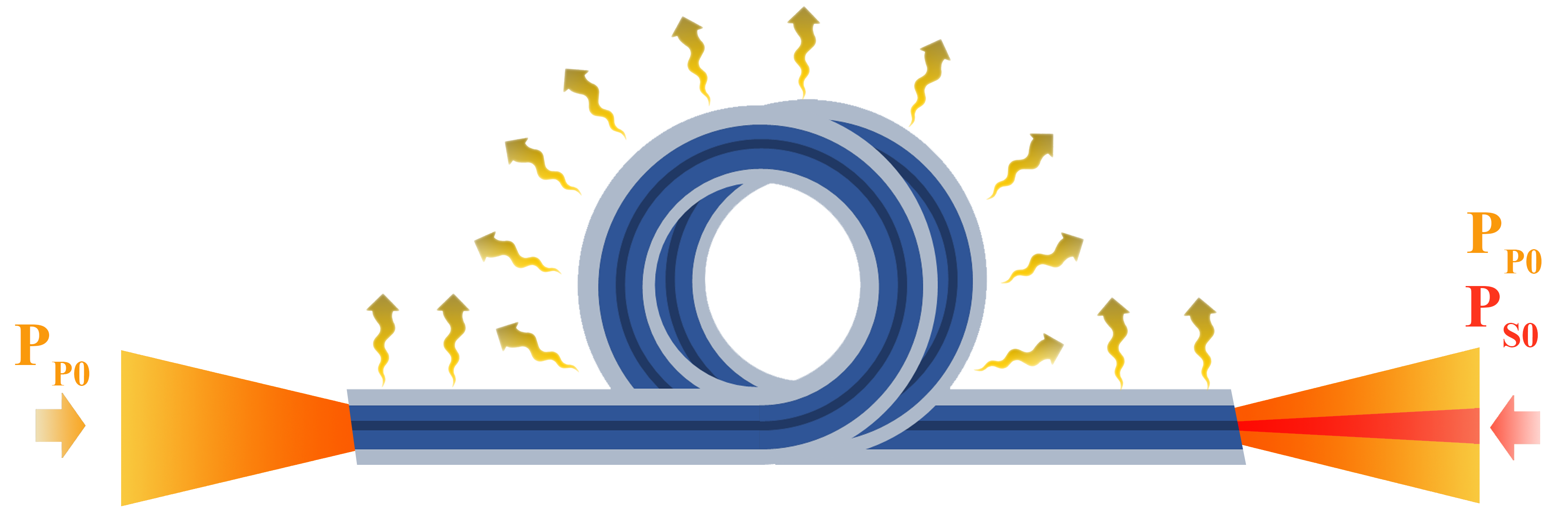}
\caption{Schematic of the suggested C/C ion-doped amplifier, $P_{p0}$ and $P_{s0}$ represent the input pump and signal seed powers.}
\label{Fig:schem}
\end{figure}

In this section, we study the suggested configuration of the C/C Yb-doped fiber amplifier which is inspired by a DC fiber amplifier described in Ref.~\cite{beier2017single}
as its schematic is shown in Fig.~\ref{Fig:schem}. For the following simulation as it was mentioned earlier, we consider two different host materials such as silica and ZBLAN glasses. 
For each host material, the contributions of the pump, signal, and fluorescence emission powers to the total LHD are calculated. Using the calculated heat density, the transverse 
and longitudinal distributions of the temperature in the amplifier are also obtained. 

As it is shown in Fig.~\ref{Fig:Cross-sectional-fiber}, the suggested configuration consists of a core and inner cladding with radii of $a$ and $b$, respectively. As it is clear in Fig.~\ref{Fig:schem}, the pump power can be coupled into the fiber amplifier from either port 1 or port 2 at $z=0$ and $z=L$, respectively. The signal power is seeded into the fiber amplifier from port 2 at $z=L$. We also 
take the fractional signal power in the core to be $\eta_{a}=0.9$. For a single-pass pump C/C Yb-doped configuration, we pump the fiber amplifier from port 1, and for a double-pass pump configuration we pump the fiber amplifier from both port 1 and port 2. Because the heat generation and extraction here are mainly governed by the pump power distribution along the fiber amplifier, we have chosen a double-pass pump C/C Yb-doped configuration where the net absorbed pump power takes a more uniform distribution along the fiber amplifier; Consequently, the temperature distribution will have a more uniform distribution along the double-pass pump fiber amplifier compared to the single-pass pump one. 

All the simulations are carried out with equally spaced signals over the emission and absorption cross-sections spaced by $\delta \lambda\,=2$~nm. Because each host material has its own emission and absorption cross-sections with a different spectral domain, the spectral domain for each ion-doped glass is different. For the C/C Yb-doped silica fiber amplifier, the considered spectral domain in the modeling starts at $\lambda_1=932$~nm ($n=1$) and ends at $\lambda_{130}=1190$~nm, which entails $n=130$ signal segments. For the Yb-doped ZBLAN glass, the spectral domain also starts at
 $\lambda_1=932$~nm (n=1) and ends at $\lambda_{80}=1090$~nm, which entails $n=80$ signal segments. 
\subsection{Simulation of a Core/Cladding Yb-doped Silica Fiber Amplifier}
 We start our modeling by considering a silica glass as the host material of the configuration with the core and inner cladding radii of $a=13~\mu m$ and $b=143~\mu m$, respectively.
 The fiber amplifier is pumped at $\lambda_{p}=1040$~nm and seeded at $\lambda_{s}=1070$~nm with the seed power of $P_{s0}^{-}=1~$W. The radiative lifetime of both core and inner 
cladding are assumed to be the same, $\tau_{r}=1$~ms. The background absorption is also taken to be {$\alpha_{b}=10$~dB/km where $\alpha_{b,a}=5$~dB/km and $\alpha_{b,s}=5$~dB/km. The mean 
fluorescence wavelength and the fiber length are $\lambda_{f}=1008$~nm and $L=26$~m, respectively~\cite{Mobini-PhysRevApplied.11.014066}. Here we assume that the Yb ion density in the core and inner cladding are $N_{01}=1 \times 10^{26}$~m$^{-3}$ and $N_{02}=6 \times 10^{25}~$m$^{-3}$, respectively. 

It is worth mentioning that with the increase in the dopant density, the internal quantum efficiency of doped-materials due to quenching process decreases. Therefore, 
one needs to pay attention to the fact that the dopant density in a host material must be smaller than the quenching concentration ~\cite{auzel2003radiation,hehlen2007model}. 
Here, we assume that a near unity internal quantum efficiency, for an ion density of $N_{02}$ in a high purity silica glass, in the presence of some modifiers like Al$_{2}$O$_{3}$, is achievable~\cite{arai1986aluminum}. However, for the sake of a higher signal amplification, we take a higher Yb ion density in the core ($N_{01}$), which comes at the cost of a lower 
internal quantum efficiency due to the fact that the ion density gets closer to the quenching concentration. Therefore, we assume that the internal quantum efficiencies of the core and 
inner cladding are $\eta_{q1}=0.8$ and $\eta_{q2}=1.0$, respectively. 

In the modeling, we consider a double-pass pump C/C Yb-doped silica fiber amplifier in which the fiber amplifier is pumped from both end facets, ports 1 and 2, at $P_{p0}^{-}=P_{p0}^{+}=0.5~$kW.
 Figure \ref{Fig:PumsignalSilica-2} shows the propagation of the right- and left- moving pump and signal powers along the fiber amplifier in the suggested double-pass pump C/C Yb-doped silica fiber amplifier. The signal power is amplified up to $P^{-}_{s}(0)=0.13\,$ KW which is equivalent to a signal efficiency of $\eta_{s}=13~\%$.
\begin{figure}[!h]
    \includegraphics[width=3.3 in]{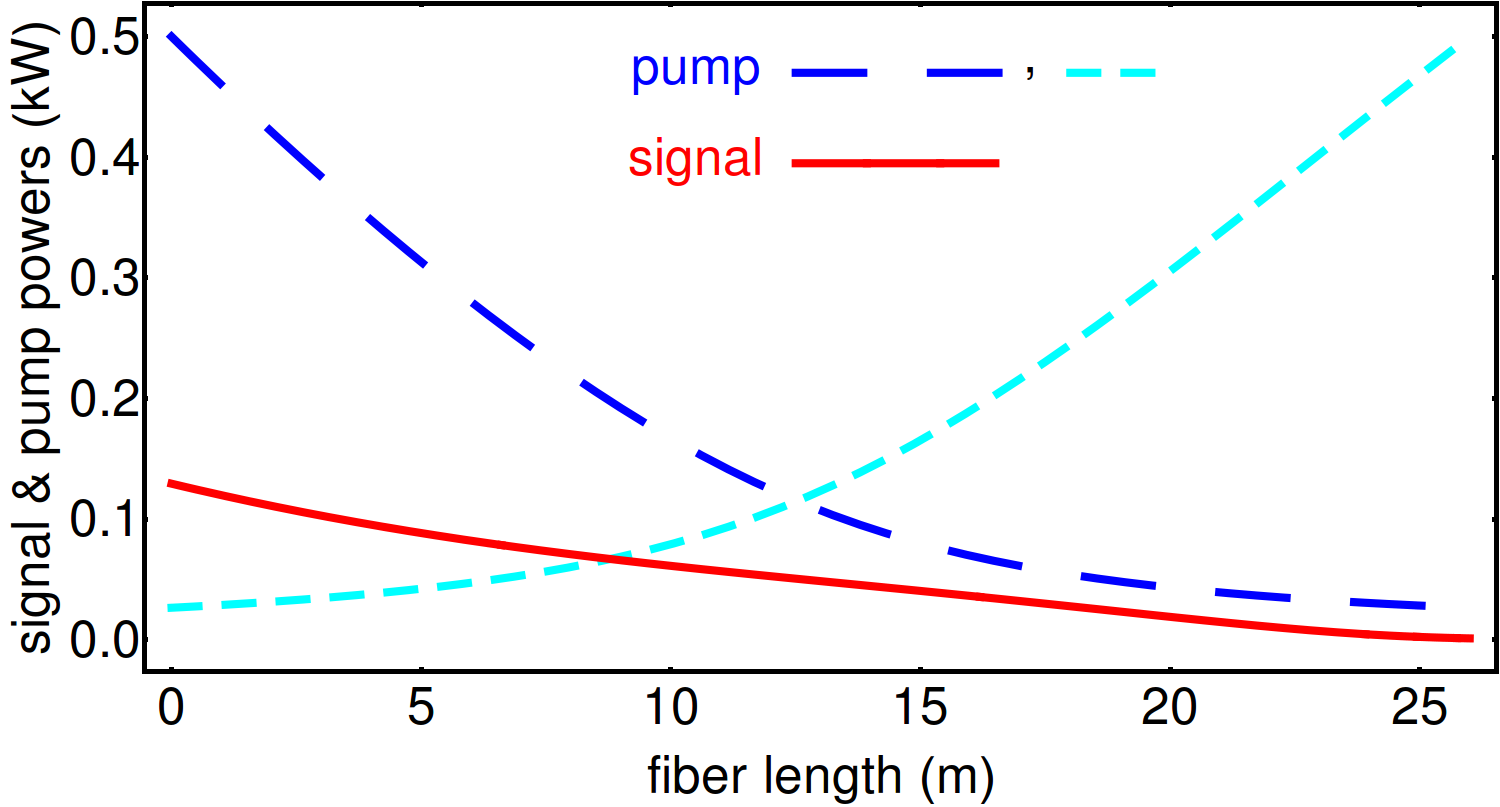}
\caption{Propagation of right- and left- moving pump and signal powers along the double-pass pump C/C Yb-doped silica fiber amplifier at $\lambda_{p}=1040$~nm and 
$\lambda_{s}=1070~$nm with $P_{p0}^{+}=P_{p0}^{-}=0.5$~kW.}
\label{Fig:PumsignalSilica-2}
\end{figure}

 In Fig.~\ref{Fig:LHDPSFout-2}, we compare the contribution of the anti-Stokes fluorescence to the net effect of the other terms that contribute to the total LHD. Here, $q^{\prime}$ ($q^{\prime}=q-q_{f}$) represents the net contributions of the quantum defect in the fiber, resonant absorption in the inner cladding plus the absorptive part of the background absorption to the 
total LHD, and $q_{f}$ represents the anti-Stokes fluorescence LHD. As it is obvious from Fig.~\ref{Fig:LHDPSFout-2}, $q^{\prime}$ and $q_{f}$ are comparable to each other; therefore, one can expect an effective heat mitigation from the anti-Stokes fluorescence  cooling in the C/C Yb-doped fiber amplifier. Fig.~\ref{Fig:LHDPSFout-2} also shows the contribution of the background absorption ($q_{b,a}$) 
to $q^{\prime}$, which is negligible compared to the other factors such as the resonant-absorption and the anti-Stokes fluorescence.
\begin{figure}[!h]
    \includegraphics[width=3.37 in]{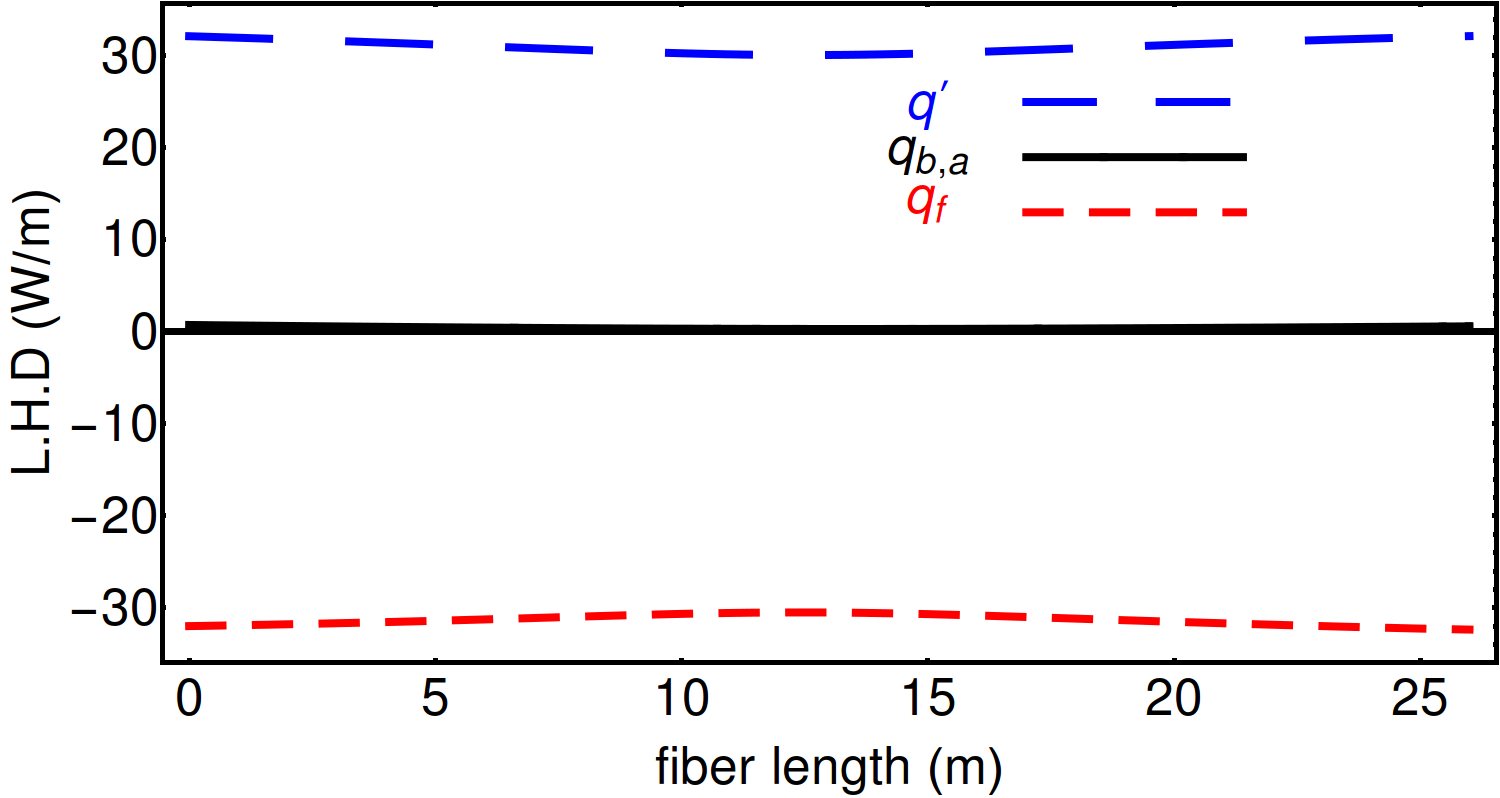}
\caption{Propagation of $q^{\prime}$, $q_{b,a}$, $q_{f}$ in the double-pass pump C/C Yb-doped silica fiber amplifier with $P_{p0}^{+}=P_{p0}^{-}=0.5\,$kW.}
\label{Fig:LHDPSFout-2}
\end{figure}

Figure \ref{Fig:Tem-along-Silica-2} describes the longitudinal distribution of the temperature along the double-pass pump C/C Yb-doped silica fiber amplifier, $\Delta T=T(0,z)-T_{0}$
, where $T_{0}$ is the ambient temperature. The temperature, a few meters after port 1, drops below the ambient temperature and retains its negative sign up till port 2. The longitudinal 
average of the temperature has also a negative value, $\overline {\Delta T}=(1/L)\int_{0}^{L}T(0,z)dz-T_{0}$, $\overline{\Delta T}=-4.2$~K for the studied configuration. 
\begin{figure}[!h]
    \includegraphics[width=3.2 in]{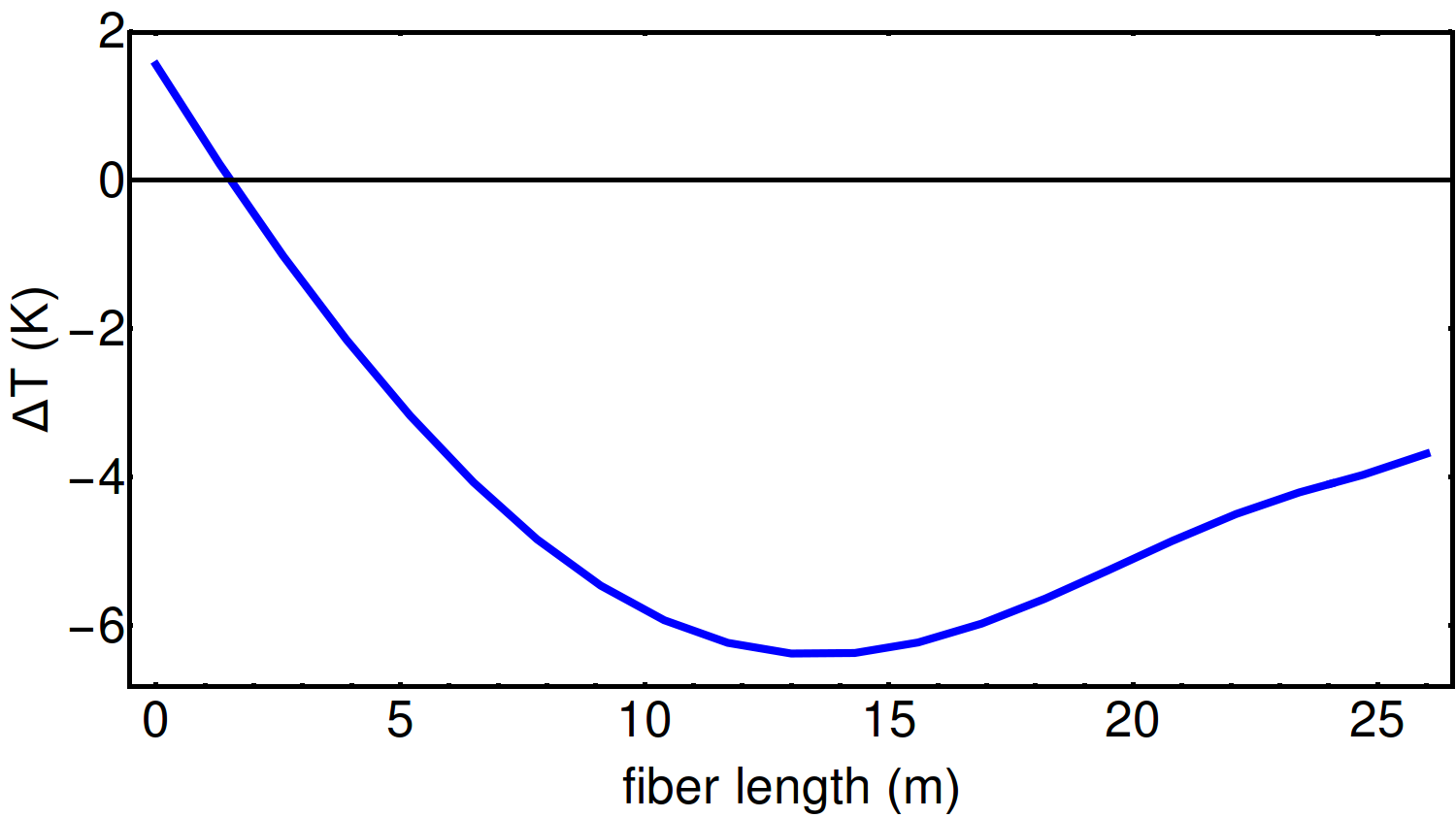}
\caption{Longitudinal temperature distribution ($\Delta T$) of the double-pass pump C/C Yb-doped silica fiber amplifier with $P_{p0}^{+}=P_{p0}^{-}=0.5\,$kW.}
\label{Fig:Tem-along-Silica-2}
\end{figure}

Figure \ref{Fig:Tem-rad-Silica-2} describes the radial distribution of the temperature in the core of the fiber amplifier at port 1. Figure \ref{Fig:Tem-rad-Silica-2} also shows 
a small temperature variation of $\delta T=T(0,0)-T(a,0)=0.02$~K across the fiber core. This low temperature variation across the core is a good indication of
 a high degree of thermalization in the C/C ion-doped configuration~\cite{Mobini:17}. 
\begin{figure}[!h]
    \includegraphics[width=3.35 in]{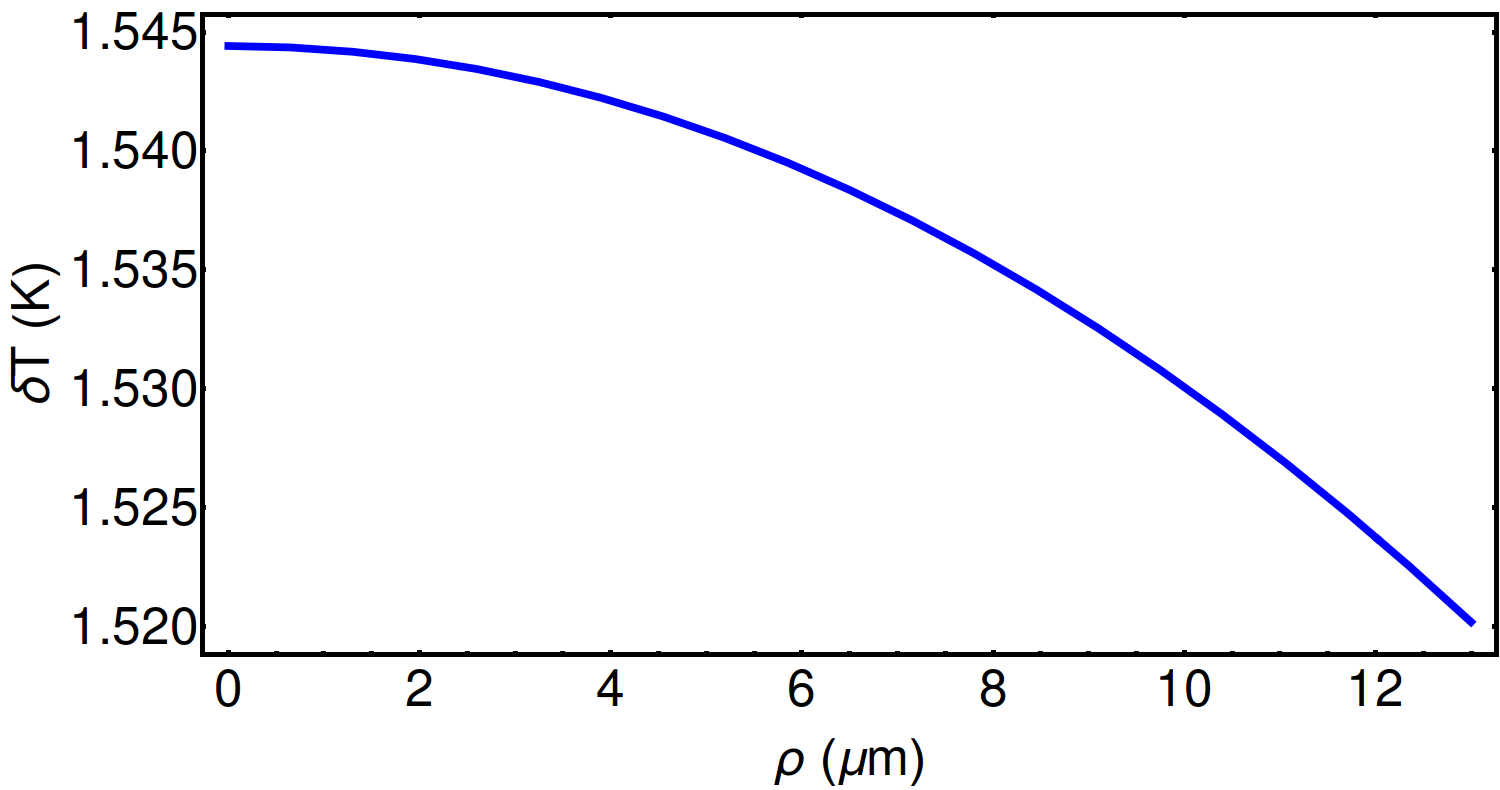}
\caption{Temperature variation across the core at port 1 for the double-pass pump C/C Yb-doped silioca fiber amplifier with $P_{p0}^{+}=P_{p0}^{-}=0.5\,$kW.}
\label{Fig:Tem-rad-Silica-2}
\end{figure}

We have shown that the average temperature of the suggested C/C Yb-doped silica fiber amplifier can go below the ambient temperature while it is delivering hundreds of Watts of signal power.
 In order to examine how effective the anti-Stokes fluorescence cooling has been in the previous example, we need to compare the previous studied case to  a typical Yb-doped DC 
fiber amplifier whose features are the same as the previous studied case except for its inner cladding, which is not doped. In other words, in the absence of the ion dopant in the inner cladding, 
the C/C ion-doped fiber amplifier turns into a typical DC fiber amplifier that lacks the anti-Stokes cooling mechanism for the heat extraction in the inner cladding.

Figure \ref{Fig:PumsignalSilica-nodop} shows the propagation of both right- and left- moving pump and signal powers along the DC fiber amplifier. The signal power is 
amplified up to $P^{-}_{s}(0)=0.23$~KW, which is equivalent to a signal efficiency of $\eta_{s}=23~\%$. The reason behind the higher signal amplification in the case compared to the 
double-pass pump C/C Yb-doped silica fiber amplifier is the absence of the Yb dopant in the inner cladding. When the inner cladding is not doped with the Yb ions, the pump power mainly 
gets absorbed in the doped core of the DC fiber amplifier and results in a higher population inversion; therefore, a higher signal amplification is obtained for the DC fiber amplifier.   
\begin{figure}[!h]
    \includegraphics[width=3.3 in]{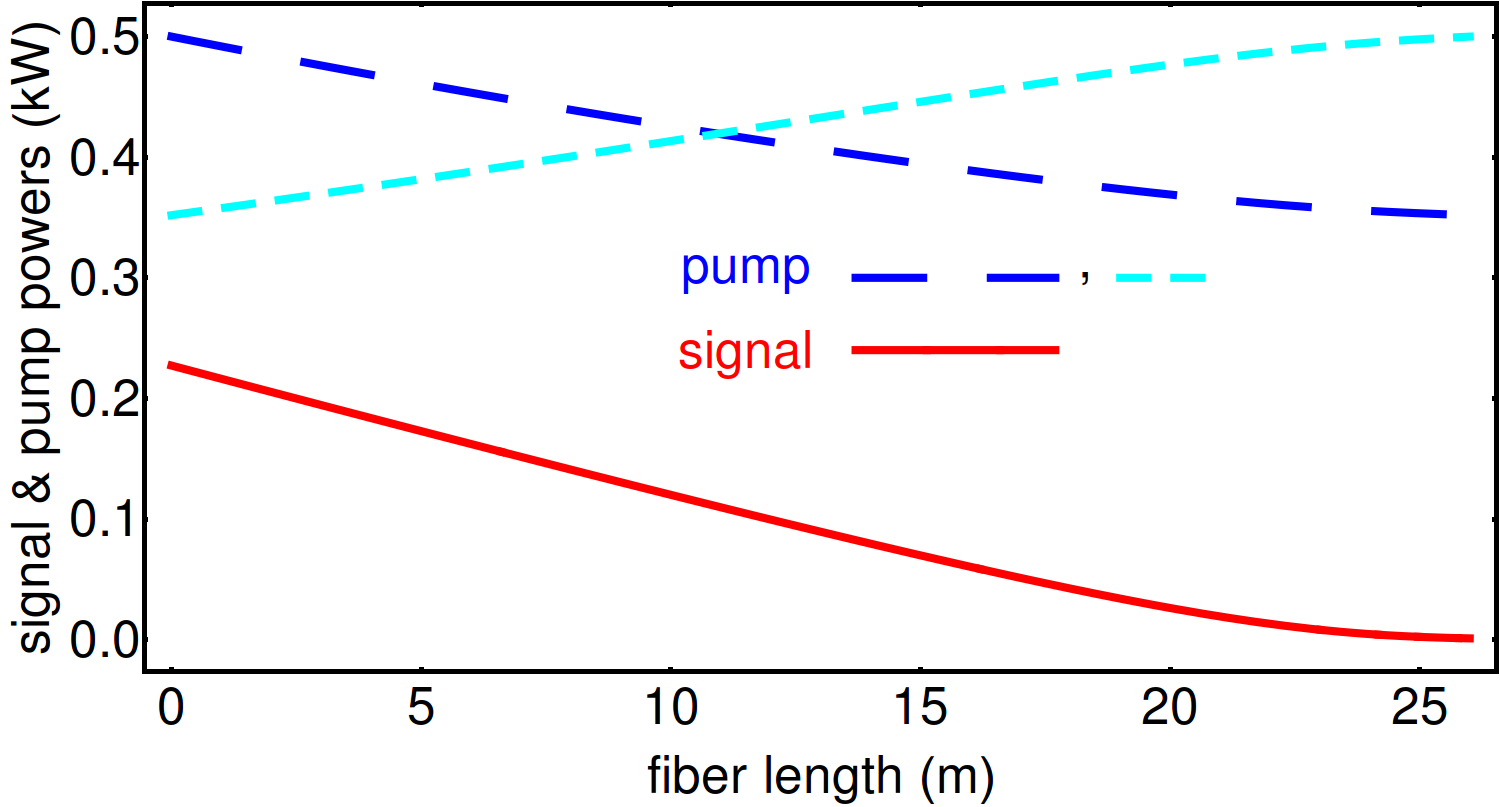}
\caption{Propagation of right- and left- moving pump and signal powers along the Yb-doped silica DC fiber amplifier at $\lambda_{p}=1040\,nm$ and $\lambda_{s}=1070\,nm$
with $P_{p0}^{+}=P_{p0}^{-}=0.5$~kW.}
\label{Fig:PumsignalSilica-nodop}
\end{figure}

Figure~\ref{Fig:LHDPSFout-nodop} represents the total LHD in the absence of the anti-Stokes cooling ($q^{\prime}$) and the anti-Stokes fluorescence LHD ($q_{f}$) in the DC fiber amplifier. $q_{f}$ is now considerably smaller than $q^{\prime}$, because the inner cladding is not doped and much of $q^{\prime}$ comes from the background absorption ($q_{ba}$) as it was explained earlier~\cite{Mobini:188}.  
\begin{figure}[!h]
    \includegraphics[width=3.4 in]{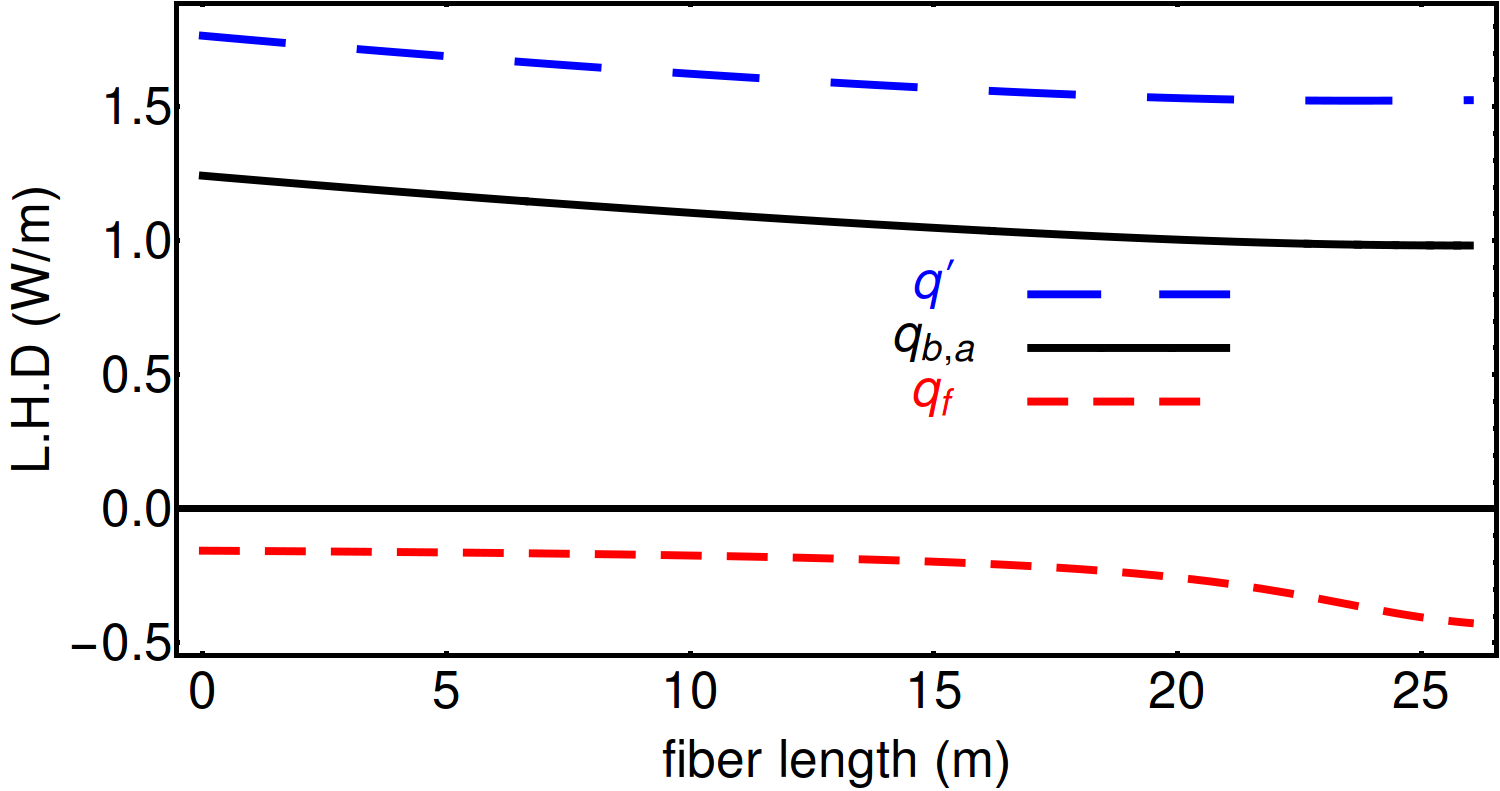}
\caption{Propagation of $q^{\prime}$, $q_{b,a}$ and $q_{f}$ in the DC Yb-doped silica fiber amplifier with $P_{p0}^{+}=P_{p0}^{-}=0.5$~kW.}
\label{Fig:LHDPSFout-nodop}
\end{figure}

Figure \ref{Fig:Tem-along-Silica-nodop} describes the longitudinal distribution of the temperature along the Yb-doped silica DC fiber amplifier. The fiber temperature is now appreciably
 higher than that of double-pass pump C/C Yb-doped silica fiber amplifier, where the anti-Stokes fluorescence cooling is responsible for the effective heat offset in the inner cladding. 
The longitudinal average of the temperature along the DC fiber amplifier is now $\overline {\Delta T}=20$~K, which is nearly $24$~K warmer than that obtained in the double-pass pump C/C 
Yb-doped configuration.  It is wroth mentioning that the temperature variation across the core of the fiber amplifier at port 1 is $\delta T=0.03$~K.
\begin{figure}[!h]
    \includegraphics[width=3.3 in]{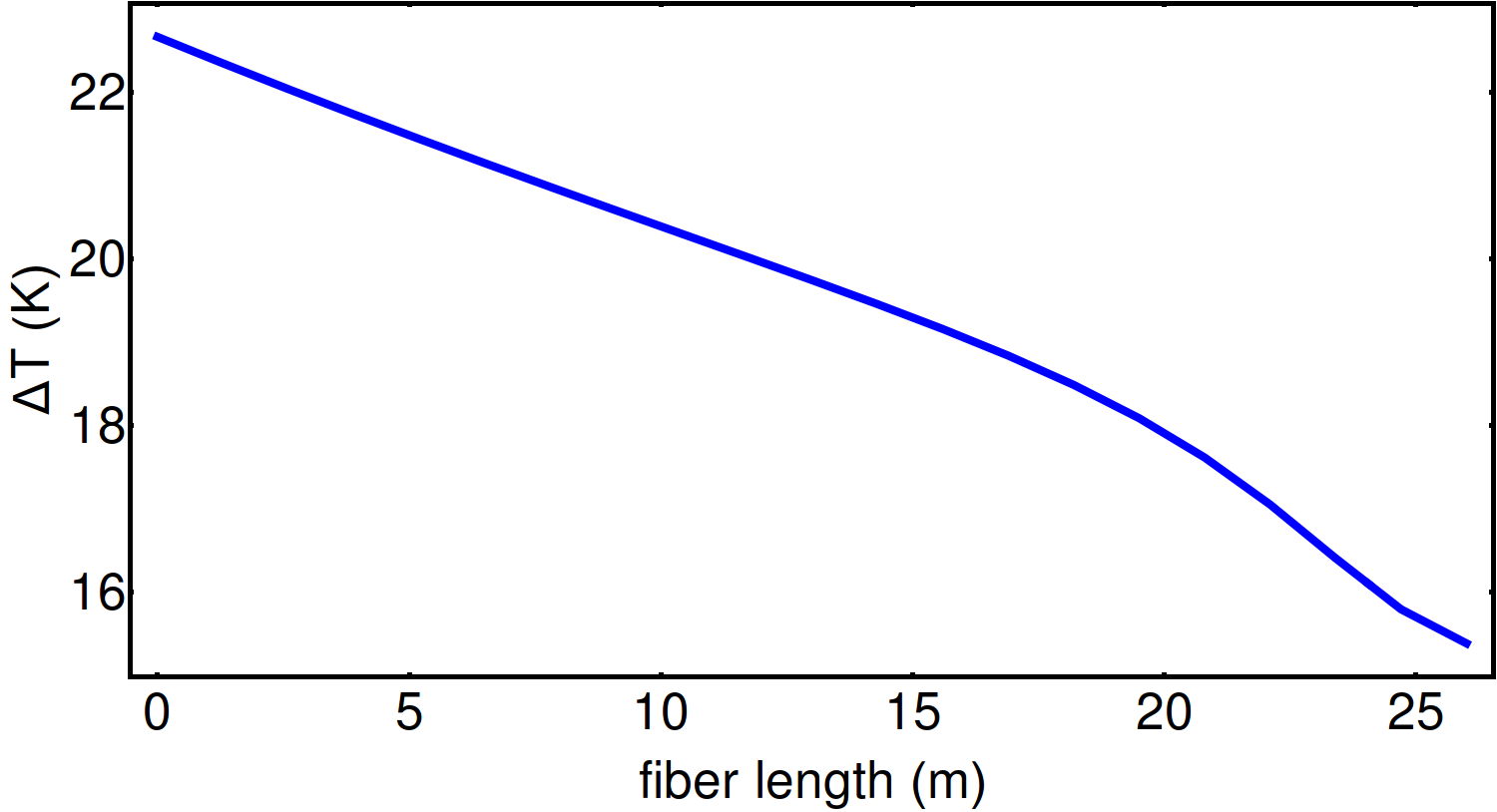}
\caption{Longitudinal temperature distribution ($\Delta T$) of the Yb-doped silica DC fiber amplifier with $P_{p0}^{+}=P_{p0}^{-}=0.5$~kW.}
\label{Fig:Tem-along-Silica-nodop}
\end{figure}

\subsection{Core/Cladding Yb-doped ZBLAN Fiber Amplifier}
As it was mentioned earlier, the ZBLAN glass as a viable host material for the solid-state laser cooling is also an interesting host-material for the C/C Yb-doped fiber amplifier. For the 
modeling, we only consider a single-pass pump C/C Yb-doped ZBLAN fiber amplifier with the core and inner cladding radii of $a~=~13~\mu m$ and $b~=~182~\mu m$~\cite{Tokita:09}. The fiber 
amplifier is pumped from port 1, at $\lambda_{p}~=~1020~$nm with an input power of $P_{p0}^{+}~=~0.5$~KW and seeded at $\lambda_{s}~=~1050$~nm with the seed power of $P_{s0}^{-}~=~1$~W. The radiative lifetimes of both core and inner cladding are assumed to be $\tau_{r}~=~1.8$~ms. The mean fluorescence wavelength and the fiber length are also taken to be $\lambda_{f}~=~995$~nm and $L~=~9$~m, 
respectively~\cite{epstein1995observation,Peysokhan:18,gosnell1999laser,gosnell1998laser}. Here, we also assume that $\alpha_{b}=20$~dB/km and $\alpha_{b,a}~=~10$~dB/km. 

One of the most important limiting factors of the ZBLAN glass is its low optical damage threshold intensity ($2.5 \times 10^{11}$~W/m$^{2}$)~\cite{zhu2010high}. 
Therefore, there is a maximum value for the signal power inside the doped core beyond which the optical damage happens. Simple calculations tell us that for the current
C/C Yb-doped ZBLAN fiber amplifier, the signal power cannot go beyond $P_{th}^{optical}=0.1$~kW~\cite{Dawson:08}. The low signal power also demands a lower  
pump power, which is equivalent to a lower heat generation in the inner cladding of the C/C Yb-doped ZBLAN fiber amplifier. Hence, to cancel out the heat generation 
that comes from the resonant and background absorptions, a lower value of the Yb dopant in the inner cladding is required. We take the Yb ion density in 
the core and the inner cladding to be $N_{01}=30\times 10^{25}$~m$^{-3}$ and $N_{02}=5 \times 10^{25}$~m$^{-3}$, respectively. The results published 
in~Refs.~\cite{gosnell1999laser,gosnell1998laser} imply that the implemented Yb ion densities of the core and inner cladding in the configuration are less than the 
quenching concentration in a pure ZBLAN glass; therefore, an internal quantum efficiency of $\eta_{q}=1$ is achievable. It is also worth mentioning that for the 
sake of a better signal amplification in the core, we have taken the Yb density of the core to be higher than that of the inner cladding. Note that the other parameters which
 are not mentioned here are the same as the parameters used in the subsection~4.A.  

Figure \ref{Fig:PumsignalZBLAN} describes the pump and signal powers along the fiber amplifier. The pump power is totally
absorbed and signal power is amplified up to $P_{s}^{-}(0)=0.06$~KW, which is equivalent to a signal efficiency of $\eta_{s}=12\%$. 
\begin{figure}[!h]
    \includegraphics[width=3.3 in]{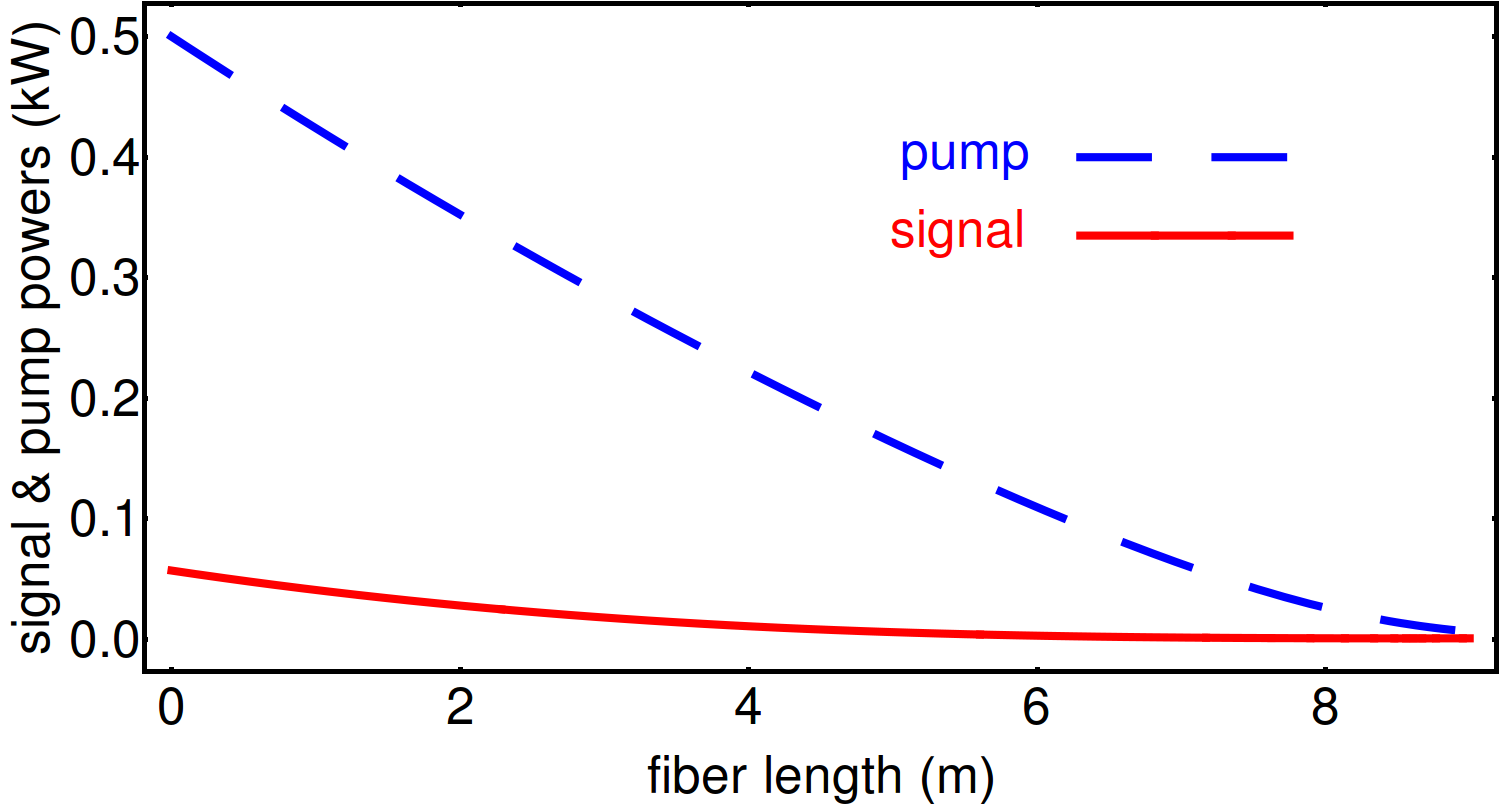}
\caption{Propagation of the pump and signal powers along the single-pass pump C/C Yb-doped ZBLAN fiber amplifier with $\lambda_{p}=1020\,nm$ and $\lambda_{s}=1050\,nm$ and
$P_{p}^{+}(0)\,=\,0.5$~kW.}
\label{Fig:PumsignalZBLAN}
\end{figure}

Figure~\ref{Fig:LHDPSFout-ZBLAN} shows the total LHD in the absence of the anti-Stokes cooling ($q^{\prime}$) and the anti-Stokes fluorescence LHD ($q_{f}$) in the C/C Yb-doped ZBLAN fiber amplifier.
$q^{\prime}$ and $q_{f}$ are comparable to each other and consequently one should expect an effective heat mitigation in the C/C Yb-doped ZBLAN configuration. 
\begin{figure}[!h]
    \includegraphics[width=3.4 in]{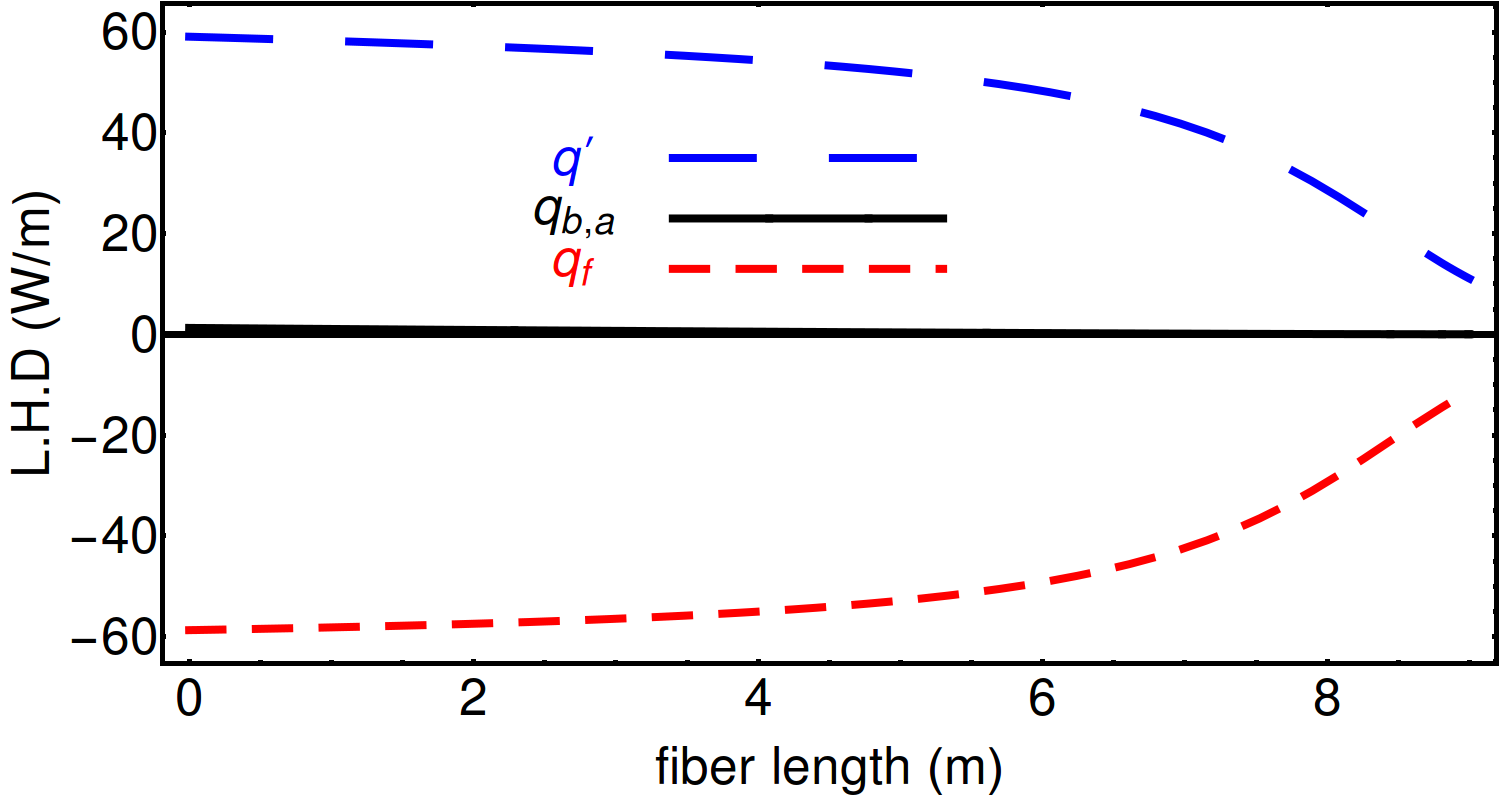}
\caption{Propagation of $q^{\prime}$, $q_{b,a}$, $q_{f}$ in the single-pass pump C/C Yb-doped ZBLAN fiber amplifier with $P_{p}^{+}(0)\,=\,0.5$~kW.}
\label{Fig:LHDPSFout-ZBLAN}
\end{figure}

Figure~\ref{Fig:Tem-along-ZBLAN} also describes the temperature distribution along the fiber amplifier. It is clear that in less than 2 meters after port 1, the amplifier temperature drops 
below the ambient temperature. The calculations show that the longitudinal average of the temperature is $\overline {\Delta T}=-5.1$~K. The obtained results also show a low temperature variation 
of $\delta T=0.08$~K at port 1 in the studied case.
\begin{figure}[!h]
    \includegraphics[width=3.4 in]{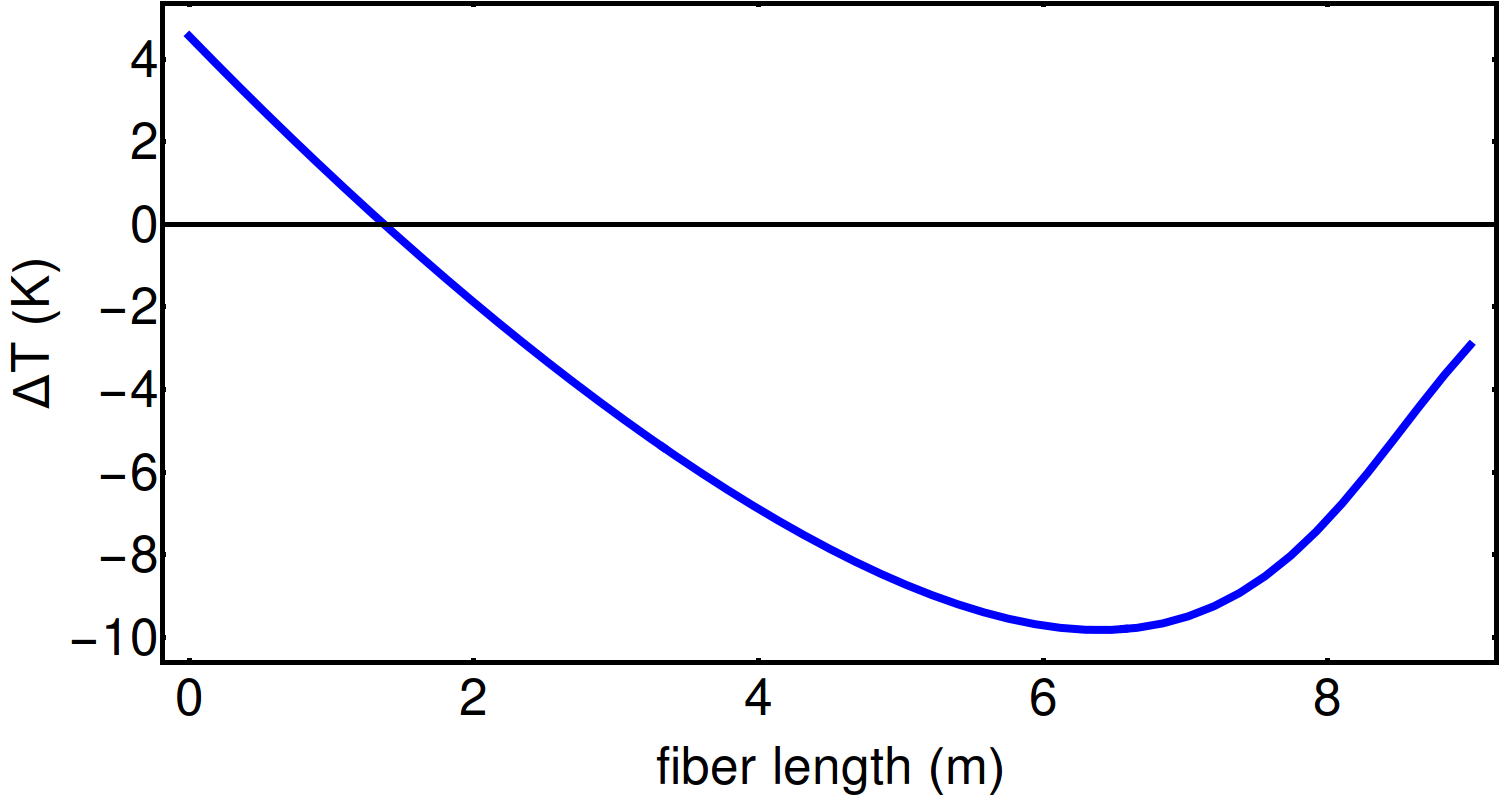}
\caption{Longitudinal temperature distribution ($\Delta T$) of the single-pass pump C/C Yb-doped ZBLAN fiber amplifier with 
$P_{p0}^{+}\,=\,0.5$~kW.}
\label{Fig:Tem-along-ZBLAN}
\end{figure}

In order to check the impact of the anti-Stokes fluorescence cooling on the heat mitigation in the studied case, we again follow the same steps and take the Yb ion density of 
the inner cladding to be $N_{02}~=~0$. For the C/C Yb-doped ZBLAN fiber amplifier in the absence of the Yb ions in the inner cladding, the calculations predict
$\eta_{s}=22\%$, and $\overline{\Delta T}=16$~K. The comparison similar to the Yb-doped silica DC fiber amplifier that was studied earlier shows an increase in both signal efficiency
 and the temperature. The increase in the average temperature is nearly $21$~K which again clearly shows that the C/C Yb-doped configuration has been effective in the heat
 mitigation for the ZBLAN glass. 
\section{Discussion}
So far we have investigated a few different examples of the suggested C/C Yb-doped fiber amplifier, where the temperature distribution, the distribution of the pump and signal powers, and the 
total LHD have been studied. In this section, we will obtain a general insight about the signal efficiency and heat loading of the C/C Yb-doped configuration. Although the main focus of the paper 
is the heat mitigation via anti-Stokes fluorescence cooling, the signal efficiency is also another important parameter for fiber amplifiers that needs to be studied. We know that in the C/C 
Yb-doped configuration, a considerable amount of the pump power is absorbed in the inner cladding due to the resonant absorption. In other words, unlike a typical DC fiber 
amplifier where the pump power mainly gets absorbed in the core, here in the C/C Yb-doped configuration, the pump power gets absorbed mainly in the inner cladding  as well as the core which results 
in a lower devoted pump power to the core; Hence, the lower pump power in the core leads to a lower signal gain or a lower signal efficiency in the C/C Yb-doped configuration.

To explore a relationship between the heat mitigation and signal efficiency in the fiber amplifier, we will base our analyses upon a few simplifications such as i) both pump and signal intensities
 are uniform across the fiber amplifier, ii) it is only the amplified signal ($\lambda_{s}$) that is taken into account in the calculations and the other emission spectra ($\lambda_{j},j\neq s$) are neglected, iii) the signal power is totally confined inside the fiber core, and iv) the fiber length is so small that the pump and signal powers do not drop appreciably along the fiber amplifier. 

Considering the above assumptions, we first start the calculations with the total LHD to obtain a simpler expression for it. Another important simplification that we make here is that we neglect the effect of the core on the calculation of the total LHD; therefore, we can reduce the whole structure of the C/C Yb-doped fiber amplifier to a Yb-doped fiber with the radius of $b$. Now, the pump power in the fiber amplifier is being attenuated by the resonant and the background absorptions. Considering all the 
simplifications and the effect of the anti-Stokes fluorescence in the modeling, the total LHD can be rewritten as follows:
\begin{align}
q(P_{p})&= \big(\frac{\zeta(\lambda_{p})}{1+p_{p}}+\alpha_{b,a}\big)P_{p},\nonumber\\
\zeta(\lambda)&=(1-\frac{\lambda}{\lambda_{f}})\sigma^{a}(\lambda)N_{02}.
\label{Eq:OptLHD}
\end{align}
\begin{figure}[!h]
    \includegraphics[width=3.3 in]{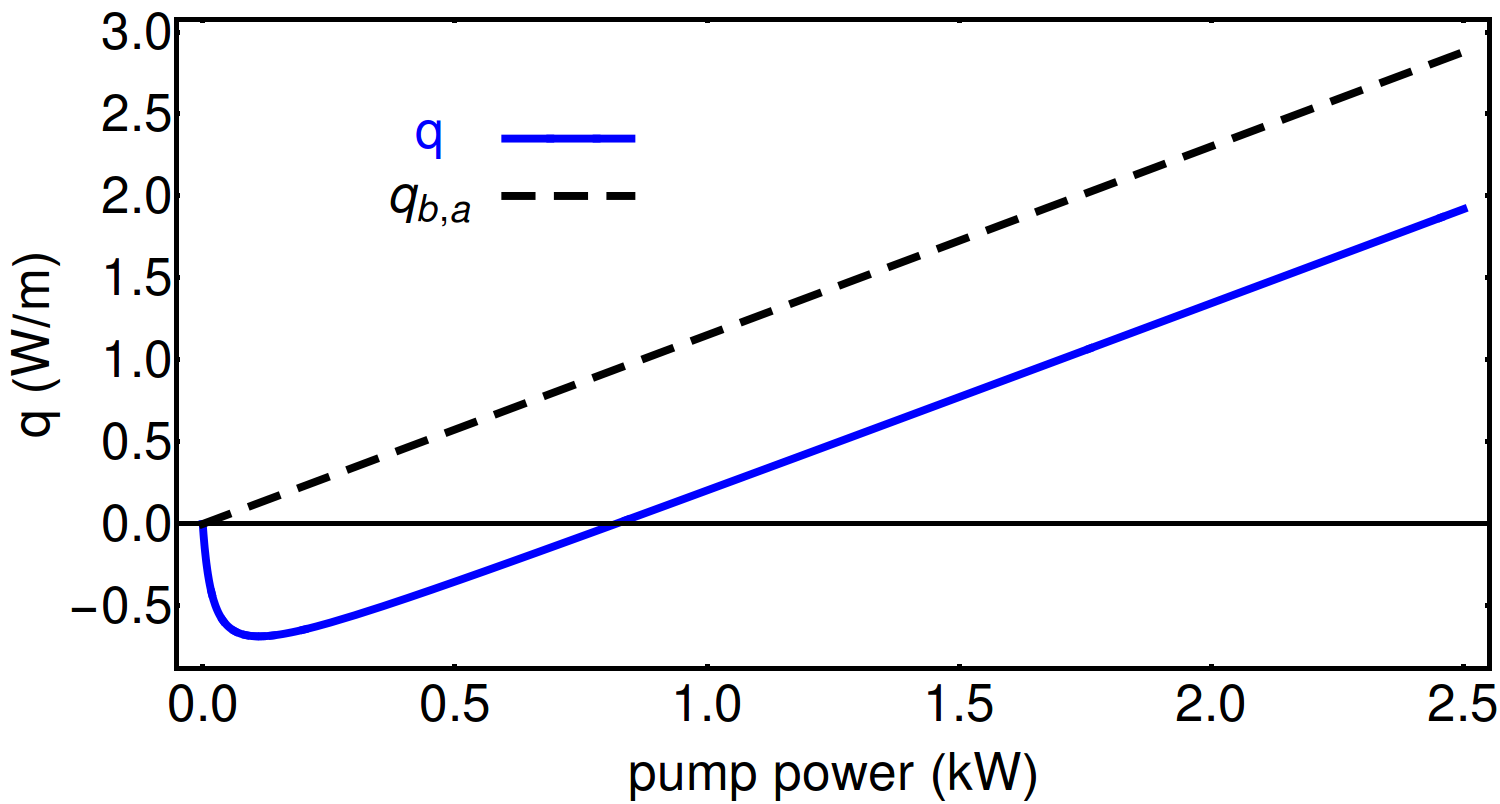}
\caption{Blue solid line describes the total LHD (q) for $\lambda_{p}=1040~$nm and $b\,=\,143~\mu m$. The cyan dashed-line described the total LHD that includes
 only the absorptive part of the background absorption ($q_{b,a}\,=\,\alpha_{b,a}P_{p}$).}
\label{Fig:OptLHD}
\end{figure}
Figure \ref{Fig:OptLHD} shows the behavior of the total LHD ($q(P_{p})$) versus the pump power for the parameters used in section~4.A where $\lambda_{p}\,=\,1040$~nm and 
$\lambda_{f}\,=\,1008$~nm. The algebra shows that the total LHD for the pump power smaller than the saturation pump power is negative or exothermic. With increasing 
the pump power, the total LHD starts decreasing until it reaches its minimum value which can be simply described by
\begin{align}
P_{p}^{opt}=P_{p}^{sat}\big(\sqrt{-\frac{\zeta(\lambda_{p})}{\alpha_{b,a}}}-1\big),
\label{Eq:optpower}
\end{align}
where for the parameters used in section~4.A, $P_{p}^{opt}=116$~W.

As the pump power goes beyond its optimum value, the resonant absorption of the pump power starts getting saturated, while the second part in Eq.~\ref{Eq:OptLHD}, $\alpha_{b,a}P_{p}$,
 that originates from the parasitic absorption linearly increases with the pump power. Therefore, the total LHD starts increasing and beyond a particular pump power becomes positive or 
endothermic as it is obvious from Fig.~\ref{Fig:OptLHD}. The maximum pump power ($P_{p}^{max}$) beyond which the total LHD becomes endothermic also can be described by   
\begin{align}
P_{p}^{max}= - P_{p}^{sat}\big({\frac{\zeta(\lambda_{p})}{\alpha_{b,a}}}+1\big).
\label{Eq:maxpower}
\end{align}
where in this case $P_{P}^{max}=0.87$~kW.

Equation \ref{Eq:maxpower} gives us a general insight about the thermal featues of the CC ion-doped fiber amplifier and an upper value for the pump power beyond which cooling
cannot be observed in the configuration. However, due to presence of the signal power in the fiber amplifier, the maximum pump power predicted in Eq.~\ref{Eq:maxpower} naturally shifts down to
the lower values as it is clear in Fig.~\ref{Fig:Tem-along-Silica-2}. Figure \ref{Fig:Tem-along-Silica-2} shows that at nearly $2\,$m from port 1, the amplifier temperature goes below 
the ambient temperature which is nearly equivalent to having a total pump power of 0.5~kW (See Fig.~\ref{Fig:PumsignalSilica-2}), so it is less than the calculated maximum power 
in Eq.~\ref{Eq:maxpower}, $P_{p}^{max}=0.87$~kW.

As the input pump power increases, the maximum pump power ($P_{p}^{max}$) also needs to shift to the higher values to guarantee more effective heat mitigation via anti-Stokes fluorescence cooling in the C/C Yb-doped configuration. This shift in the maximum pump power can be done either by increasing the dopant density in the inner cladding or increasing the inner 
cladding area. Because the quenching concentration is a barrier to the dopant density increase in the configuration, an increase in the inner cladding area is a better choice.  
Eq.~\ref{Eq:maxpower} shows that with the increase of the inner cladding area, the maximum pump power increases linearly. Therefore, a C/C Yb-doped configuration with a larger inner cladding 
area is more desirable as long as the heat mitigation via anti-Stokes fluorescence cooling is concerned.

So far we have learned that with the increase of the pump power, the inner cladding area needs to be increased for more effective heat mitigation. But beside the heat mitigation for the
 C/C Yb-doped configuration, one also needs to investigate the signal efficiency ($\eta_{s}$) of the C/C Yb-doped configuration. To get a better insight about the signal efficiency of the 
C/C Yb-doped configuration, we introduce a quantity based on the previous simplifications as follows (Here we also have neglected the background absorption):
\begin{align}
\label{Eq:materialeff}
\eta_{s}&= \frac{dP_{s}}{dz}/\frac{dP_{p}}{dz},\\
\frac{1}{\eta_{s}}&=\frac{1}{\eta_{s}^{DC}}+\nonumber(1-\Gamma_{p})\frac{P_{p}}{P_{s}}\frac{N_{02}}{N_{01}}\frac{\sigma_{p}^{a}}{\sigma_{s}^{a}}\frac{(1+\frac{p_{s}}{1+p_{p}})}{(\frac{\beta_{p}}{\beta_{s}}-1)p_{p}-1}\nonumber,
\end{align}
where $\eta_{s}^{DC}$ represents the signal efficiency of a DC fiber amplifier whose parameters are the same as the C/C ion-doped fiber amplifier except for the ion dopant in the 
inner cladding where $N_{20}=0$.

From Eq.~\ref{Eq:materialeff}, one realizes that the signal efficiency of the C/C ion-doped fiber amplifier is lower than a typical DC fiber amplifier
 due to the resonant absorption that takes place in the inner cladding. Another important point that can issue from Eq.~\ref{Eq:materialeff} is that for a constant core diameter, as the inner 
cladding diameter increases, which is equivalent to the decrease of the pump overlapping factor ($\Gamma_{p}$), the signal efficiency decreases. Therefore, as long as the signal efficiency
 is concerned, a smaller inner cladding diameter is more desirable which is in sharp contrast with what we have learned from Eqs.~\ref{Eq:OptLHD}-\ref{Eq:maxpower}, where a larger cross-section 
of the inner cladding is more desirable for more effective heat mitigation. Hence, this important incompatibility should be taken into account in the design of the 
C/C ion-doped configuration.
\section{conclusion}
We have presented a new design of a DC fiber amplifier in which the inner cladding of the DC fiber is doped with the same ion as in the core.
By doping the inner cladding of a typical DC fiber amplifier, we create a situation in which the anti-Stokes fluorescence cooling can offset the generated heat 
from the background absorption effectively in the inner cladding . In the modeling, we presented a detailed analytical formalism that not only considers the 
spatial profiles of both pump and signal intensities, but also entails ASE as the source of the anti-Stokes fluorescence cooling. Using the obtained signal and pump powers
 along and across the C/C Yb-doped fiber amplifier, the generated heat and consequently temperature distribution in the the C/C Yb-doped amplifier were calculated. 
In the study, two host glasses amenable to the solid-state laser cooling such as silica and ZBLAN glasses are considered. The calculations show 
that as long as the internal quantum efficiencies of the host materials are very close to unity, the anti-Stokes fluorescence cooling can decrease the excess heat effectively 
 in the C/C Yb-doped configuration in high-power operation.
\section{Appendix}
\subsection{Propagation of pump power}
Calculations of the pump power propagation in the core and inner cladding can be described by 
\begin{align}
\label{Eq:pump-intensity}
\pm \frac{dP^{\pm}_{p}}{dz}&=-\frac{1}{\pi b^2}\Big(\sigma_{p}^{a} N_{01} P_{p}^{\pm}\int_{0}^{2 \pi}d\phi\int_{0}^{a}\frac{1+g_{w}(\rho)\sum\limits_{j=1}^n \gamma_{j} \tilde{p}_{j}^{\pm}}{1+\tilde{p}_{p}^{\pm}+g_{w}(\rho)\sum\limits_{j=1}^n\tilde{p}_{j}^{\pm}} \rho d\rho \nonumber\\
&-\alpha_{b} P^{\pm}_{p}\int_{0}^{2 \pi}d\phi\int_{0}^{a}\rho d\rho\nonumber\\
&+\sigma_{p}^{a} N_{02} P_{p}^{\pm}\int_{0}^{2 \pi}d\phi\int_{a}^{b}\frac{1+g_{w}(\rho)\sum\limits_{j=1}^n \gamma_{j} \tilde{p}_{j}^{\pm}}{1+\tilde{p}_{p}^{\pm}+g_{w}(\rho)\sum\limits_{j=1}^n\tilde{p}_{j}^{\pm}} \rho d\rho \nonumber\\
&-\alpha_{b}P^{\pm}_{p}\int_{0}^{2 \pi}d\phi\int_{a}^{b}\rho d\rho\Big).
\end{align}
To calculate the above terms, we have used the following identity:
\begin{align}
&\int_{a}^{b} \frac{A+B g_{w}(\rho)}{C+D g_{w}(\rho)} \rho d\rho=\frac{A}{C} \frac{(b^2-a^2)}{2}+\frac{w^2}{4}(\frac{AD-BC}{CD})\nonumber \\
&\times \ln\big[\frac{C+(1-\eta_{b})D}{C+(1-\eta_{a})D}\big].
\label{Eq:pump-power2}
\end{align}
\subsection{Propagation of signal power}
Calculations of the signal power propagation in the core and inner cladding can be described by
\begin{align}
\pm \frac{dP^{\pm}_{j}}{dz}&=\int_{0}^{2 \pi}d\phi\int_{0}^{a}\frac{\beta_{p}\tilde{p}_{p}^{\pm}+g_{w}(\rho)\sum\limits_{k=1}^n \beta_{k} \tilde{p}_{k}^{\pm}}{1+\tilde{p}_{p}^{\pm}+g_{w}(\rho)\sum\limits_{k=1}^n \tilde{p}_{k}^{\pm}} N_{01}f_{w}(\rho)\rho d\rho \nonumber\\
&\times\Big((\sigma_{j}^{a}+\sigma_{j}^{e})P_{j}^{\pm}+\sigma_{j}^{e}\Pi_{j}\Big)\nonumber\\
&-N_{01} \sigma_{j}^{a}P_{j}^{\pm}\int_{0}^{2 \pi}d\phi\int_{0}^{a}f_{w}(\rho)\rho d\rho\nonumber\\
&-\alpha_{b} P^{\pm}_{j}\int_{0}^{2 \pi}d\phi\int_{0}^{a}f_{w}(\rho)\rho d\rho \nonumber\\
&+\int_{0}^{2 \pi}d\phi\int_{a}^{b}\frac{\beta_{p}\tilde{p}_{p}^{\pm}+g_{\omega}(\rho)\sum\limits_{k=1}^n \beta_{k} \tilde{p}_{k}^{\pm}}{1+\tilde{p}_{p}^{\pm}+g_{w}(\rho)\sum\limits_{k=1}^n \tilde{p}_{k}^{\pm}} N_{02}f_{w}(\rho)\rho d\rho \nonumber\\
&\times\Big((\sigma_{j}^{a}+\sigma_{j}^{e})P_{j}^{\pm}+\sigma_{j}^{e}\Pi_{j}\Big)\nonumber\\
&-N_{02} \sigma_{j}^{a}P_{j}^{\pm}\int_{0}^{2 \pi}d\phi\int_{a}^{b}f_{w}(\rho)\rho d\rho\nonumber\\
&-\alpha_{b} P^{\pm}_{j} \int_{0}^{2 \pi}d\phi\int_{a}^{b}f_{w}(\rho)\rho d\rho.
\label{Eq:signal-power1}
\end{align}

To calculate the above terms, we have used the following identities:
\begin{align}
&\int_{a}^{b} \frac{A+B g_{w}(\rho)}{C+D g_{w}(\rho)} f_{w}(\rho) \rho d\rho=\nonumber\\
&\frac{B}{D}(\eta_{b}-\eta_{a})-\big(\frac{AD-BC}{D^2}\big)
\times \ln\big(\frac{C+D(1-\eta_{b})}{C+D(1-\eta_{a})}\big)\Big), 
\label{Eq:Integral2}
\end{align}
\begin{align}
\int_{0}^{2 \pi}d\phi\int_{x_{0}}^{x_{1}}f_{w}(\rho)\rho d\rho=\eta_{x_{1}}-\eta_{x_{0}}.
\label{Eq:signal-power-4stterm}
\end{align}
\subsection{Anti-Stokes Fluorescence}
Calculations of the anti-Stokes fluorescence along the fiber amplifier in the core and inner cladding can be described by
\begin{align}
&\frac{dP_{f}}{dz}=\frac{hc}{\lambda_{f}\tau_{r}}\Big(N_{01}\int_{0}^{2 \pi}d\phi\int_{0}^{a}\frac{\beta_{p}\tilde{p}_{p}^{\pm}+g_{\omega}(\rho)\sum\limits_{j=1}^n \beta_{j} \tilde{p}_{j}^{\pm}}{1+\tilde{p}_{p}^{\pm}+g_{\omega}(\rho)\sum\limits_{j=1}^n \tilde{p}_{j}^{\pm}}\rho d\rho \nonumber\\
&+ N_{02}\int_{0}^{2 \pi}d\phi\int_{a}^{b}\frac{\beta_{p}\tilde{p}_{p}^{\pm}+g_{\omega}(\rho)\sum\limits_{j=1}^n \beta_{j} \tilde{p}_{j}^{\pm}}{1+\tilde{p}_{p}^{\pm}+g_{\omega}(\rho)\sum\limits_{j=1}^n \tilde{p}_{j}^{\pm}}\rho d\rho \Big).
\label{Eq:FluorEm}
\end{align}
\subsection{Temperature distribution across the fiber amplifier}
The temperature distribution inside the core and inner cladding of the fiber amplifier can be described by 
\begin{align}
&T_{co}(\rho)=\frac{a^2}{4 k}\big(Q_{co}(\tilde{P}_p^{\pm}, \tilde{P}_j^{\pm}) - Q_{inc}(\tilde{P}_p^{\pm}, \tilde{P}_j^{\pm})\big)\nonumber\\
&+\frac{a^2}{2 b H_c}(Q_{co}(\tilde{P}_p^{\pm}, \tilde{P}_j^{\pm}) - Q_{inc}(\tilde{P}_p^{\pm}, \tilde{P}_j^{\pm}))\nonumber\\
&+\frac{b^2}{4k}Q_{inc}(\tilde{P}_p^{\pm}, \tilde{P}_j^{\pm})+\frac{b}{2H_c} Q_{inc}(\tilde{P}_p^{\pm}, \tilde{P}_j^{\pm})-\frac{\rho^2}{4 k} Q_{co}(\tilde{P}_p^{\pm}, \tilde{P}_j^{\pm})\nonumber\\
&+\frac{a^2}{2 k} (Q_{co}(\tilde{P}_p^{\pm}, \tilde{P}_j^{\pm})-Q_{inc}(\tilde{P}_p^{\pm}, \tilde{P}_j^{\pm})) \ln(\frac{b}{a})+T_0,
\label{Eq:Tempin}
\end{align}
and
\begin{align}
&T_{inc}(\rho)=\frac{a^2}{2 b H_c} Q_{co}(\tilde{P}_p^{\pm}, \tilde{P}_j^{\pm})+\frac{b^2}{4 k} Q_{inc}(\tilde{P}_p^{\pm}, \tilde{P}_j^{\pm})\nonumber\\
 &+\frac{b^2-a^2}{2 b H_c} Q_{inc}(\tilde{P}_p^{\pm}, \tilde{P}_j^{\pm})-\frac{\rho^2}{4 k} Q_{inc}(\tilde{P}_p^{\pm}, \tilde{P}_j^{\pm})\nonumber\\
 &-\frac{a^2}{2k}(Q_{co}(\tilde{P}_p^{\pm}, \tilde{P}_j^{\pm})-Q_{inc}(\tilde{P}_p^{\pm}, \tilde{P}_j^{\pm}))\times \ln(\frac{\rho}{b})+T_0,
\label{Eq:Tempout}
\end{align}
 where $T_{0}=300$~K, is the ambient temperature, $H_{c}\,=\,80~W/m^2K$ is the heat convective coefficient and $k$ is the thermal connectivity of the host material ($k=1.23$ for silica glass and $k=0.63$ for ZBLAN glass).
\section{Funding Information}
This material is based upon work supported by the Air Force Office of Scientific Research under award number FA9550-16-1-0362
titled Multidisciplinary Approaches to Radiation Balanced Lasers (MARBLE).

\end{document}